\documentclass[12pt]{revtex4-1}
\usepackage{graphicx}
\usepackage{amsmath,amssymb}
\newcommand{\fl}{}
\newcommand{\pd}[1]{\partial_{#1}}

\newcommand{\bnabla}{\mbox{\boldmath $\nabla$}}
\renewcommand{\vec}[1]{\mbox{\boldmath $#1$}}

\usepackage[usenames,dvipsnames]{color}
\newcommand{\APW}[1]{\textcolor{black}{#1}}
\newcommand{\AG}[1]{\textcolor{black}{#1}}

\graphicspath{{./small_fig/}}

\begin{document}

\newpage
\title[]{Transition to magnetorotational turbulence in Taylor--Couette flow with imposed azimuthal magnetic field}

\author{A. Guseva$^1$, A. P. Willis$^2$, R. Hollerbach$^3$, M. Avila$^1$}

\address{$^1$ Friedrich-Alexander University Erlangen-N\"urnberg 
anna.guseva@fau.de}
\address{$^2$ School of Mathematics and Statistics, University of Sheffield, Sheffield S3 7RH, UK}
\address{$^3$ School of Mathematics, University of Leeds, Leeds LS2 9JT, UK}

\vspace{10pt}
%
\begin{abstract}
The magnetorotational instability (MRI) is thought to be a powerful source of turbulence and momentum transport in astrophysical accretion discs, but obtaining observational evidence of its operation is challenging. Recently, laboratory experiments of Taylor--Couette flow with externally imposed axial and azimuthal magnetic fields have revealed the kinematic and dynamic properties of the MRI close to the instability onset. While good agreement was found with linear stability analyses, little is known about the transition to turbulence and transport properties of the MRI.  We here report on a numerical investigation of the MRI with an imposed azimuthal magnetic field. We show that the laminar Taylor--Couette flow becomes unstable to a wave rotating in the azimuthal direction and standing in the axial direction via a supercritical Hopf bifurcation. Subsequently, the flow features a catastrophic transition to spatio-temporal defects which is mediated by a subcritical subharmonic Hopf bifurcation. Our results are in qualitative agreement with the PROMISE experiment and dramatically extend their realizable parameter range. We find that as the Reynolds number increases defects accumulate and grow into turbulence, yet the momentum transport scales weakly.
\end{abstract}

%
%
%
\maketitle
%
%

\section{Introduction}

The magnetorotational instability (MRI) is of great importance in astrophysics. First discovered by Velikhov~\cite{velikhov1959stability} in 1959, it remained unnoticed until 1991 when Balbus and Hawley  \cite{balbus1991} realised its application to accretion disc theory. Accretion discs are astrophysical systems that consist of ionised gas and dust orbiting a massive body. Planets and stars are formed from this initially dispersed matter. The physical mechanism of accretion is straightforward: a parcel of viscous fluid in the differentially rotating disc loses its angular momentum over time and falls onto the central object. To explain the astrophysically observed rates of accretion, however, one must assume a turbulent transport of angular momentum in the outward direction \cite{shakura1973}. In so-called Keplerian discs the angular velocity profile of gas  follows the law
\begin{equation}
\label{eq:keplerian}
\Omega\sim r^{-3/2},
\end{equation}
which is hydrodynamically stable according to the Rayleigh criterion for  rotating fluids \cite{rayleigh1917dynamics}:
 \begin{equation}
 \label{eq:rayleigh}
 \frac{d(r^2\Omega)^2}{dr}>0 \quad\text{for stability}.
 \end{equation}
Ionized accretion discs, however, are necessarily magnetized and the MRI may still act in rotating flows, provided the angular velocity decreases with radius, which is true of Keplerian flows (\ref{eq:keplerian}).

The growth rates of the MRI and the parameter ranges in which it acts were determined in several linear analyses~\cite{balbus1991,ogilvie1996non,balbus1998instability,hollerbach2005new,hollerbach2010nonaxisymmetric}, but these do not provide information about the flow structure and scaling of angular momentum transport  after nonlinear saturation. In the last two decades there has been a great deal of numerical work concerned with the nonlinear properties of the MRI. \AG{Simulations are usually performed with the shearing sheet approximation, which is a local model of an accretion disc with shear-periodic boundary conditions in the radial direction~\cite{balbus1998instability}. The main disadvantage of this model is the influence of boundary conditions on the geometry of the observed modes and transport scaling. In particular, the length of the computational box fixes the modes that appear and determines their nonlinear saturation. As it is not clear how the length should be selected, the interpretation of the transport scaling becomes quite involved~\cite{umurhan2007weakly}.}

\begin{figure}
  \begin{center}
    \begin{tabular}{cc}
     (a) \hspace{2cm} & (b) \\ \includegraphics[width=0.28\linewidth]{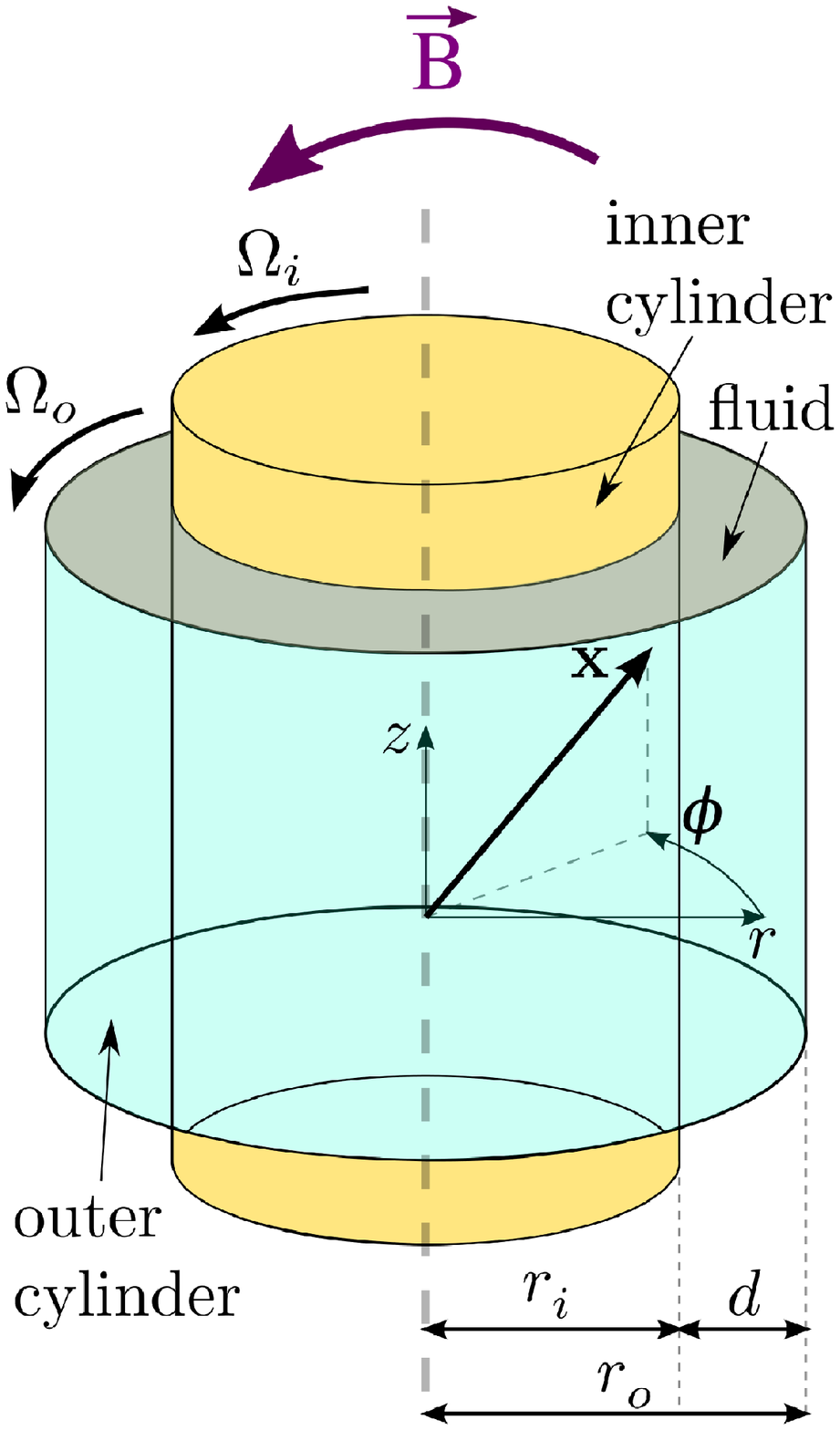} \hspace{1cm} & 
    \includegraphics[width=0.6\linewidth]{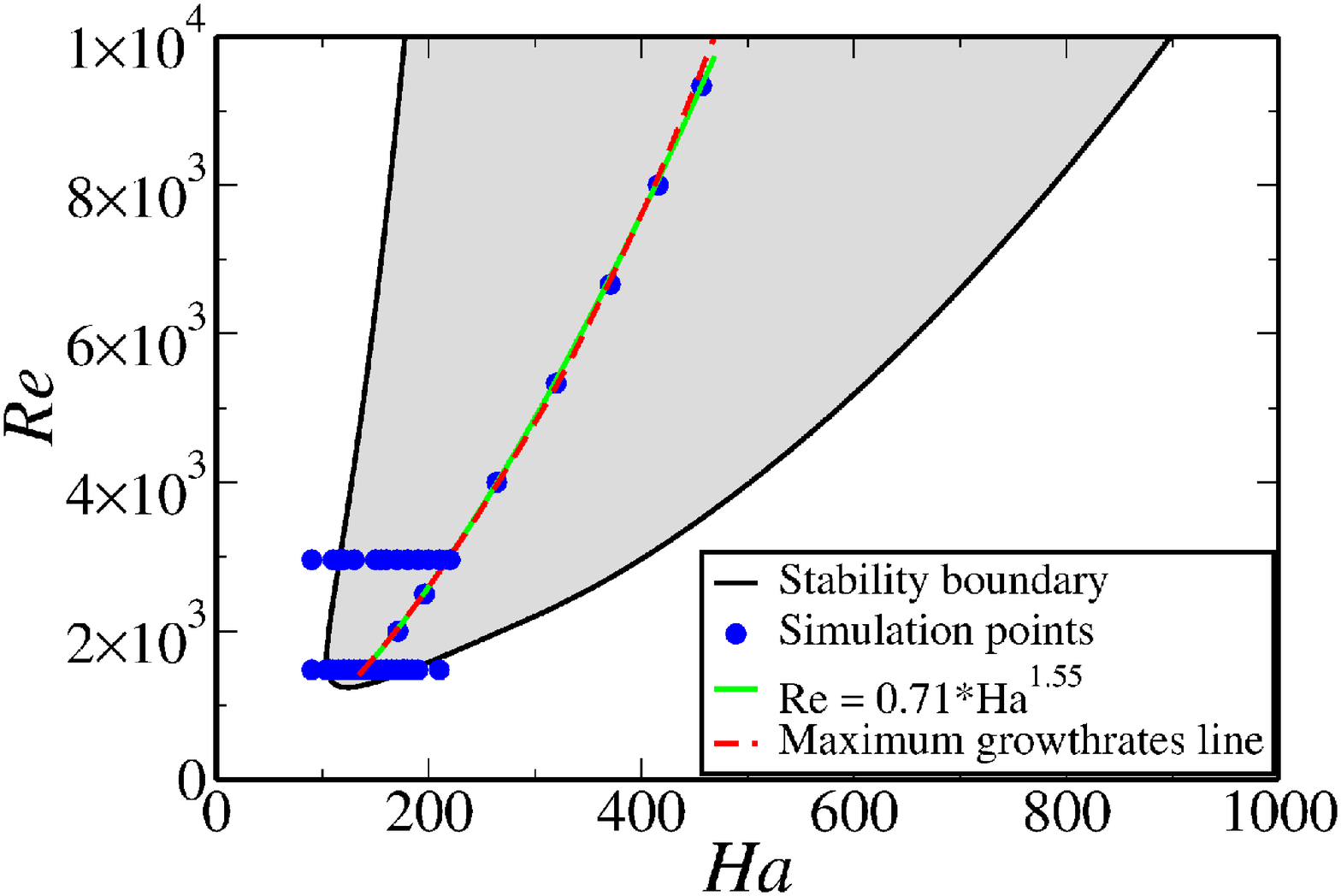}\\
    \end{tabular}
  \end{center}
  \caption{\small ($a$) \textbf{Schematic of the Taylor-Couette geometry with azimuthal magnetic field}. A  liquid metal is confined between two coaxial cylinders of radii $r_i$ and $r_o$, which can rotate independently at angular velocities $\Omega_i$ and $\Omega_o$. In this work the rotation rate is fixed  at $\mu=\Omega_o/\Omega_i=0.26$ and an azimuthal magnetic field of the form $(r_i/r)B_0$ is imposed, where $r$ is the radial coordinate. ($b$) \textbf{Instability region of the AMRI}.  \AG{Blue circles correspond to the points at which our simulations were conducted, the red dashed line to the curve of maximum growth rate and the green solid line is a fit to the latter of the form $Ha=a Re^b$, where $a=0.71$ and $b=1.55$.}} 
  \label{fig:TC_scheme}
\end{figure}

These theoretical results and numerical simulations inspired physicists to
realise the MRI in laboratory experiments. In 2001 Ji et al.~\cite{ji2001magnetorotational} and R{\"u}diger and Zhang~\cite{rudiger2001mhd} independently suggested the possibility of directly observing magnetorotational instabilities in a cylindrical vessel made of two co-axial and independently rotating cylinders containing a liquid metal alloy (see Fig. \ref{fig:TC_scheme}a). The standard form of the MRI (SMRI), which they proposed, emerges when a purely axial magnetic field is imposed, but this has not yet been achieved in experiments. The difficulty is that liquid metals have very small magnetic Prandtl numbers (e.g.\ $Pm \sim 10^{-6}$ for gallium alloys), leading in this case to very large Reynolds numbers ($Re\gtrsim 10^7$) necessary to observe the SMRI \AG{(see Table~\ref{tab:param} for the definitions of $Re$ and $Pm$)}. In fact, such high Reynolds numbers have never been achieved even for non-magnetic Taylor-Couette flows. \AG{A further difficulty of Taylor--Couette experiments in the quasi-Keplerian regime arises because of Ekman vortices that arise adjacent to the endplates. Unless a very specific endplate arrangement is used, the Ekman vortices extend deep into the flow and even at moderate Reynolds number the basic Couette flow cannot be obtained experimentally~\cite{edlund2014nonlinear}. The resulting velocity profiles are no longer quasi-Keplerian and hydrodynamic instabilities render the flow turbulent even in the absence of magnetic fields~\cite{avila2012stability,nordsiek2014azimuthal}.}

Hollerbach and R\"udiger \cite{hollerbach2005new} proposed instead a combination of axial and azimuthal magnetic fields, giving rise to the helical MRI (HMRI) at much lower $Re\sim 10^3$ for Hartmann numbers $Ha \sim 10$ \AG{(see Table \ref{tab:param} for the definition of Ha)}. This was successfully observed  \cite{stefani2006experimental,stefani2009helical} in the PROMISE facility (Potsdam-ROssendorf Magnetic Instability Experiment). Both the SMRI and HMRI consist of  axisymmetric toroidal vortices, which are stationary for the former but travel axially for the latter. Hollerbach \emph{et al}.~\cite{hollerbach2010nonaxisymmetric} realised that  although  a purely azimuthal magnetic field does not yield any axisymmetric instabilities, non-axisymmetric modes can be destabilised. The resulting azimuthal MRI (AMRI) arises in Taylor-Couette flow at $Re\sim 10^3$ for $Ha \sim 10^2$ \cite{hollerbach2010nonaxisymmetric}, and a recent upgrade of the PROMISE power supply made its experimental observation possible \cite{seilmayer2014experimental}. \AG{In  PROMISE  the endplates are split into two parts and the inner (outer) one is attached to the inner (outer) cylinder. Although this configuration is acceptable at $Re \leq 3000$, as studied experimentally, endplate effects become dominant at larger $Re$. Because of this, and of the practical impossibility of generating a purely azimuthal magnetic field experimentally, it is challenging to actually identify the AMRI modes in the experimental data unambiguously (see \cite{seilmayer2014experimental}).}

Despite this recent experimental progress in realising magnetorotational instabilities in the laboratory, little is known about their  bifurcation scenario, transition to turbulence and transport properties as $Re$ increases. In this work we address these points for the AMRI. We perform direct numerical simulations of the coupled induction and Navier-Stokes equations using axially periodic boundary conditions, thereby avoiding undesired endplate effects \AG{and focusing on the features intrinsic to the AMRI}. We find that the laminar quasi-Keplerian flow becomes unstable to a wave rotating in the azimuthal direction and standing in the axial direction. Subsequently, we identify a new bifurcation scenario giving rise to spatio-temporal defects via a subcritical subharmonic Hopf bifurcation. As the Reynolds number is further increased, the flow becomes turbulent and outward momentum transport is enhanced, albeit at a weak rate. The results are in good qualitative agreement with the PROMISE observations~\cite{seilmayer2014experimental} and substantially extend the parameter range 
explored experimentally. 

\section{Governing equations}


We consider an incompressible viscous liquid metal that is sheared between two independently rotating cylinders of radii $r_i$ (inner) and $r_o$ (outer). The angular velocity of the cylinders are $\Omega_i$ and $\Omega_o$, respectively, and an external azimuthal magnetic field $(r_i/r)B_0$, where $r$ is the radial coordinate, is imposed. The relevant fluid properties are the electrical conductivity $\sigma$, the kinematic  viscosity $\nu$ , the density $\rho$  and  the  magnetic diffusivity $\eta$. The velocity field $\mathbf{u}$ is determined by the Navier-Stokes equations (\ref{eq:NSeq}), whereas the magnetic field $\mathbf{B}$ is determined by the induction equation (\ref{eq:Ieq}), which represents a combination of the laws of Ampere, Faraday, and Ohm. The equations were rendered dimensionless by using the gap between cylinders $d=r_o-r_i$ for length, $d^2/\nu$ for time, and $B_0$ for the magnetic field. In dimensionless form they read
\begin{equation}\label{eq:NSeq}
   (\pd{t} + \mathbf{v}\cdot\nabla) \mathbf{v} = -\nabla p + \nabla^2 \mathbf{v} + \frac{Ha^2}{Pm} (\nabla \times \mathbf{B}) \times \mathbf{B},
\end{equation}
\begin{equation}\label{eq:Ieq}
( \pd{t} - \frac{1}{Pm} \nabla^2 ) \mathbf{B}  = \nabla \times (\mathbf{v} \times \mathbf{B}),
\end{equation}
together with $\nabla\cdot{\mathbf{v}}=\nabla\cdot{\mathbf{B}}=0$. Here $p$ is the pressure, $Ha$ the Hartmann number, and $Pm$ the magnetic Prandtl number. The dimensionless parameters of the system are specified in Table~\ref{tab:param}.  Following the PROMISE experiment~\cite{seilmayer2014experimental}, we use a magnetic Prandtl number of $Pm=1.4\cdot 10^{-6}$ (corresponding to the alloy $Ga^{67}In^{20.5}Sn^{12.5}$), a radius-ratio of $\delta=0.5$ and a rotation-ratio of $\mu=0.26$. This places the velocity profile in the quasi-Keplerian regime, for which the angular velocity decreases radially, whereas the angular momentum increases, i.e.~$\delta^2<\mu<1$.  

\begin{table}[!h]
\renewcommand{\arraystretch}{1.3}
  \centering
 \caption{\small \textbf{Dimensionless parameters of the magnetohydrodynamic
    Taylor-Couette problem}.}
  \begin{tabular}{|l|l|l|l|}
    \hline
    Abbrev.   & Parameter                      &Definition & Range \\
    \hline
    $\delta$ & Radius ratio                   &$r_i/r_o$ & 0.5\\
    $\alpha$ &  Axial wavenumber (geometrical parameter)      & $2\pi / L_z$ & $0.5$---$4.5$ \\
    $\mu$ &  Angular velocity ratio         &$\Omega_o/\Omega_i $ & 0.26 \\
    $Pm$ &      Magnetic Prandtl number        &$\nu/\eta$ & $1.4 \cdot 10^{-6}$ \\
    $Re$ &  Reynolds numbers of inner cylinder  &$\Omega_i r_i d/\nu$ & $1480$---$9333$\\
    $Ha$ &      Hartmann number                &$B_0 d/(\frac{\sigma}{\rho\nu})^{1/2}$ & $90$---$457$\\
    \hline
  \end{tabular}  \label{tab:param}
\end{table}

\subsection{Boundary conditions}

We employ cylindrical coordinates $(r,\phi,z) \in [r_i,r_o] \times [0,2\pi]\times [0,L_z]$, for which the no-slip velocity boundary conditions at the cylinders read
\begin{equation}
u_\phi(r_i,\phi,z)=Re, \qquad
u_\phi(r_o,\phi,z)=\mu Re.
\end{equation}
Periodicity in the axial direction is imposed with basic length $L_z$. 
The background circular Couette flow $\mathbf{V} = V(r)\mathbf{e_\phi}$ is a solution to the equations and boundary conditions given by
\begin{equation}\label{eq:couette}
   V(r) = \frac1{1+\delta}
   \left[
      ( \frac{\mu}{\delta} Re - \delta \, Re) r + \frac{\delta}{(1-\delta)^2}
      ( Re - \mu \, Re) \frac1{r}
   \right] .
\end{equation}

The magnetic field is also assumed to be periodic in the axial direction. In the radial direction the boundary condition depends on the material of the cylinders. Typically two idealized cases are considered in the MRI problem: insulating and conducting cylinders. These lead to slightly different results, as theoretically demonstrated by Chandrasekhar \cite{chandrasekhar1961}. However, the difference is not great, and here we will consider only the case of insulating boundaries.
\AG{Assuming the Fourier expansion for each variable
\begin{equation}\label{eq:fourier}
A=\sum_{|k|<K} \sum_{|m|<M} A_{k,m}(r)\exp[\mathrm{i}(\alpha kz + m\phi)],
\end{equation}
one obtains the following boundary conditions for $\vec{B}$:\\
For case $k=m=0$:
\begin{equation}
     B_\theta = B_z = 0 .
\end{equation}
case $k=0$, $m\ne 0$:
\begin{equation}
   B_r\pm\mathrm{i} B_\theta=0,\quad B_z=0
   \qquad \mbox{($+$ on $r_i$, $-$ on $r_o$)}.
\end{equation}
case $k\ne0$:
\begin{equation}
			B_r + \mathrm{i}\,\frac{\mathcal{B}'_m(\alpha k R)}
       {\mathcal{B}_m(\alpha k R)} \, B_z = 0,
      \quad
      \alpha k B_\theta - \frac{m}{r} B_z = 0,
\end{equation}
where $\mathcal{B}_m(x)$ denotes the modified Bessel function 
$I_m(x)$ for $r=r_i$ and $K_m(x)$ for $r=r_o$, and
$\mathcal{B}'_m=\pd{x}\mathcal{B}_m$. See Willis and Barenghi \cite{willis2002hydromagnetic} for a detailed derivation. } 

\subsection{Brief remarks on symmetries of rotating magnetohydrodynamic flows}\label{sec:sym}

The basic circular Couette flow  \eqref{eq:couette}  has SO(2) $\times$ O(2) symmetry, where SO(2) represents the rotational symmetry in the azimuthal direction. In the axial direction the group O(2) may be written as O(2)$=Z_2 \rtimes$SO(2), where $Z_2$ is a reflection (up-down symmetry) and SO(2) the translational symmetry in the $z$ direction. The presence of purely axial or purely azimuthal imposed magnetic field does not change the symmetry group of the system. Hence if the primary instability is a Hopf bifurcation the resulting states can be either standing or traveling waves in the axial direction \cite{crawford1991symmetry}. By contrast, a combined helical magnetic field breaks the reflection symmetry and only traveling waves (TW) can be observed \cite{knobloch1996symmetry}. Finally, if the bifurcating solution is non-axisymmetric, as in the AMRI, this will generically be a rotating wave in the azimuthal direction.

\section{Numerical method}

In the numerical simulations only the deviation from the basic flow $\mathbf{u} = \mathbf{v} - \mathbf{V}$ is computed. Its governing equations read
\begin{equation}
   (\pd{t} - \nabla^2)\, \mathbf{u} 
   \,=\, \mathbf{N} - \nabla p , \qquad
   \nabla \cdot \mathbf{u} = 0,
\end{equation}
which are supplemented with homogeneous boundary conditions $\mathbf{u} = \mathbf{0}$.
Here $\mathbf{N}$ stands for the nonlinear term in the Navier-Stokes equations 
(\ref{eq:NSeq}), which contains the advective terms and the Lorentz force:
\begin{eqnarray}
  \fl \mathbf{N} & = & \mathbf{u} \times (\nabla \times \mathbf{u}) 
   - (\mathbf{V}\cdot\nabla) \mathbf{u} - (\mathbf{u}\cdot\nabla) \mathbf{V}  + \frac{Q}{Pm} (\nabla \times \mathbf{B}) \times \mathbf{B}  \\
\fl   & = & \mathbf{u} \times (\nabla \times \mathbf{u}) 
   - (V/r) \pd{\phi} \mathbf{u}
   + (2V/r) u_\phi \mathbf{e_r}
   - u_r ( 1+\pd{r}) V \mathbf{e_\phi}
 + \frac{Q}{Pm} (\nabla \times \mathbf{B}) \times \mathbf{B}. \nonumber
   \label{eq:nonlinvel}
\end{eqnarray}
Spatial discretisation is accomplished via the Fourier expansion in the azimuthal and axial directions (\ref{eq:fourier}), and
as the variables are real, their Fourier coefficients satisfy the property  $A_{k,m}=A^*_{-k,-m}$,
where $A^*$ denotes the complex conjugate. 

\APW{The pseudospectral Fourier method is the most efficient choice for periodic boundary conditions. Because of their great accuracy at low resolutions, spectral methods have also been used to discretise the hydromagnetic Taylor--Couette flow problem in the radial direction~\cite{hollerbach2008spectral,willis2002hydromagnetic}. In these works the Chebyshev collocation method was chosen because of its simplicity. Nevertheless its computational and storage costs scale as $\mathcal{O}(N^2)$, where $N$ is the number of radial points, making computations at large Reynolds number impractical. Petrov--Galerkin formulations reduce the cost to $\mathcal{O}(N)$ and have been also used in hydrodynamic Taylor--Couette flows~\cite{Moser_jcp1983,Meseguer_EPJ2007}. However, the treatment of the radial boundary conditions becomes very cumbersome for the magnetic field.  We here use the finite-difference method in the radial direction. Radial derivatives are calculated using 9-point stencils to $(9-j)^\mathrm{th}$ order, where $j$ is the order of the derivative. This results in banded matrices with associated $\mathcal{O}(N)$ cost, while providing excellent accuracy. Ref.~\cite{LiRaHoAv14} provides a  more thorough discussion of these computational issues and the accuracy of the finite-difference method applied to  hydrodynamic Taylor--Couette flow.
} 

\APW{
A second-order scheme is applied to the time discretisation 
$t^q=q\, \Delta t$, based on the implicit Crank--Nicolson method.
Applying this method to the Navier-Stokes equations, we find
\begin{equation}\label{eq:NS-CN}
 \left(1/\Delta t - \nabla^2\right)\mathbf{u}^{q+1}=
 \left(1/\Delta t + \nabla^2\right)\mathbf{u}^q + \mathbf{N}^{q+\frac{1}{2}}
 - \nabla p, 
 \qquad
 \nabla^2 p = \nabla \cdot \mathbf{N}^{q+\frac{1}{2}},
\end{equation}
where $\mathbf{N}^{q+\frac{1}{2}}$ is an estimate for the 
nonlinear terms (Euler predictor, Crank-Nicolson corrector).
In cylindrical coordinates the $r$- and $\phi$-components of the 
Laplacian operator couples the two components, and
for a Fourier decomposition they are complex operators.  
Programming complexity and computational cost
of inversion for $u_r^{q+1}$ and $u_\phi^{q+1}$ can be reduced,
}
however, by considering
\begin{equation}
   u_\pm = u_r \pm \mathrm{i} \, u_\phi,
   \qquad \mbox{i.e.}~~
   u_r = \frac{1}{2} ( u_+ + u_-),
   \quad
   u_\phi = -\,\frac{\mathrm{i}}{2}(u_+ - u_- ) ,   
\end{equation}
where the $\pm$ are taken respectively. 
\APW{The equations governing these components separate
and the Laplacian operator is now real ($\partial_\phi\to\mathrm{i}m$),}
\begin{equation}
   (\pd{t} - \nabla^2_\pm)\, u_\pm
    =  N_\pm - (\nabla p)_\pm , \qquad
   \nabla^2_\pm = \nabla^2 - \frac{1}{r^2} 
   \pm \frac{\mathrm{i}}{r^2}\pd{\phi}.  
\end{equation}

\subsection{Influence-matrix method}

\APW{
The natural approach to solving (\ref{eq:NS-CN}) is to invert 
for $p$ and then for $\mathbf{u}^{q+1}$.  All the boundary 
conditions are on $\mathbf{u}^{q+1}$, however, there are none on $p$, 
and it
is well known that primitive variable formulations are subject 
to loss of temporal order if inappropriate boundary conditions 
are enforced on $p$ \cite{rempfer2006}.  
}
For the magnetic
field, there appear at first sight to be too few boundary conditions,
and further, the components of $\vec{B}$ are coupled 
in the boundary condition.
It is shown here that the influence-matrix method resolves these 
issues, the appropriate boundary conditions can be satisfied to 
machine precision, and temporal order is retained.  We show first
how the method is applied for time integration of the velocity field,
%
similar to \cite{marcus1984simulation}.
An analogous approach is applied to the magnetic field.

\subsubsection{Method for the velocity field}

We write the time-discretised Navier-Stokes equations 
(\ref{eq:NS-CN})
in the form
\begin{equation}
   \label{eq:NSdisc}
	 \left\{\begin{array}{rcl}
    X \mathbf{u}^{q+1} & = & Y \mathbf{u}^q 
		 + \mathbf{N}^{q+\frac{1}{2}} - \bnabla p \, , \\
		 \nabla^2 p & = & \bnabla\cdot(Y \mathbf{u}^q 
		 + \mathbf{N}^{q+\frac{1}{2}}) ,
	 \end{array}\right.
\end{equation}
where $q$ denotes time $t_q$. This form is sixth order in $r$ for 
$\mathbf{u}^{q+1}$ and second order for $p$, without the 
solenoidal condition explicitly imposed. 
In principle this system should be inverted simultaneously for $p$ and $\mathbf{u}^{q+1}$ with boundary conditions
$\mathbf{u}^{q+1}=\mathbf{0}$ and $\bnabla\cdot\mathbf{u}^{q+1}=0$.
In practice it
would be preferable to invert for $p$ first then for $\mathbf{u}^{q+1}$,
but the boundary conditions do not involve $p$ directly.  
Note that the $Y \mathbf{u}^q$ term has been included in the 
right-hand side of the pressure-Poisson equation, so that
it corresponds precisely to the divergence of the equation
for $\mathbf{u}^{q+1}$.  This ensures that any non-zero
divergence in the initial condition is projected out
after a single time-step. We split the system (\ref{eq:NSdisc}) 
into the `bulk' solution, $\{\bar{\mathbf{u}},\bar{p}\}$,
\begin{equation}
   \label{eq:NSbulk}
	 \left\{\begin{array}{rcl}
     X \bar{\mathbf{u}} & = & Y \mathbf{u}^q 
		 + \mathbf{N}^{q+\frac{1}{2}} - \bnabla \bar{p} \, , \\
		  \nabla^2 \bar{p} & = & \bnabla\cdot(Y \mathbf{u}^q 
		 + \mathbf{N}^{q+\frac{1}{2}}) ,
	 \end{array}\right.
\end{equation}
with boundary conditions $\bar{\mathbf{u}}=\mathbf{0}$ and $\pd{r}\bar{p}=0$,
and the auxiliary systems
\begin{equation}
   \label{eq:NSp0}
	 \left\{\begin{array}{rcl}
     X \mathbf{u}' & = & -\nabla p' \, , \\
		  \nabla^2 p' & = & 0,
	 \end{array}\right.
\end{equation}
with two sets of boundary conditions, $\mathbf{u}'=\mathbf{0}$ with $\pd{r}p'=\{0,1\}$ and $\{1,0\}$ on $r=\{r_i,r_o\}$, and
\begin{equation}
   \label{eq:NSu0}
	 \left\{\begin{array}{rcl}
     X \mathbf{u}' & = & \mathbf{0},
	 \end{array}\right.
\end{equation}
with boundary conditions $u'_+=\{0,1\}$, $u'_-=\{0,1\}$, 
$u'_z=\{0,\mathrm{i}\}$ and similarly their reversed versions
on $\{r_i,r_o\}$.  
\APW{
These dashed functions may be precomputed, and will be used to 
correct the approximate boundary conditions used to calculate
$\bar{\mathbf{u}}$ and $\bar{p}=0$.}

The system (\ref{eq:NSp0}) provides, 
with the two boundary conditions, two linearly independent functions 
$\mathbf{u}'_j$ that may be added to $\bar{\mathbf{u}}$ without altering the 
right-hand side in (\ref{eq:NSbulk}).  Similarly the system (\ref{eq:NSu0})
provides a further six functions.  The superposition
\begin{equation}
   \label{eq:usuperpos}
   \mathbf{u}^{q+1} = \bar{\mathbf{u}} + \sum_{j=1}^8 a_j\, \mathbf{u}'_j \, 
\end{equation}
may be formed in order to satisfy the eight original 
boundary conditions, $\mathbf{u}^{q+1}=\mathbf{0}$ and 
$\bnabla\cdot\mathbf{u}^{q+1}=0$ on $r=\{r_i,r_o\}$.
Substituting (\ref{eq:usuperpos}) into the 
boundary conditions, they may be written
\begin{equation}
   A \mathbf{a} = -\mathbf{g}(\bar{\mathbf{u}}) ,
\end{equation}
where $A=A(\mathbf{g}(\mathbf{u}'))$ 
is an 8$\times$8 matrix.  The appropriate coefficients required
to satisfy the boundary conditions are thereby recovered from the
solution of this small system for $\mathbf{a}$.
The error in the boundary conditions $g_j(\mathbf{u}^{q+1})$
using the influence-matrix technique
is at the level of the machine precision.

The auxiliary functions $u'_j(r)$, the matrix 
$A$ and its inverse may all be precomputed,
and the boundary conditions for $\mathbf{u}'$ have been chosen
so that $u'_\pm$ are purely real, 
$u'_z$ is purely imaginary, and $A$ is real.  
For each timestep, this application of the influence matrix 
technique requires only evaluation of the deviation from 
the boundary condition, 
multiplication by an 8$\times$8 real matrix,
and the addition of four functions to each component of $\mathbf{u}$, 
each either purely real or purely imaginary.
Compared to the evaluation of nonlinear terms, 
the computational overhead is negligible.

\subsubsection{Method for the magnetic field}

Consider the induction equation (\ref{eq:Ieq}) 
time-stepped without $\nabla\cdot{\mathbf{B}}=0$ 
enforced.  Evolution of $\psi=\nabla \cdot \mathbf{B}$ is 
then governed by the divergence of (\ref{eq:Ieq}),
\begin{equation}
   \label{eq:diveqn}
   \pd{t}\psi = \frac{1}{Pm}\nabla^2\psi\, .
\end{equation}
In addition to the boundary conditions derived in
\cite{willis2002hydromagnetic},
the condition $\psi=0$ must be satisfied on the boundary
to avoid introduction of divergence into the domain.
Then (\ref{eq:Ieq}) has the appropriate number of 
boundary conditions, 
\APW{and $\psi=\nabla\cdot{\mathbf{B}}$ should remain zero
for a solenoidal initial condition.}

To prevent accumulation of divergence from artificial internal sources,
i.e.\ discretisation error, it is commonplace to introduce
an artificial pressure $\Pi$ \cite{Brackbill1980effect}.
The discretised system is then as in 
(\ref{eq:NSdisc}) where one reads $\mathbf{B}$ for $\mathbf{u}$ and 
$\Pi$ for $p$.
The boundary condition 
for $\Pi$ is any choice such that, 
when one computes $\Pi$ for a given $\mathbf{B}^q$,
it is found to be constant when $\bnabla\cdot\mathbf{B}^q=0$
\cite{ramshaw1983method}.  The choice $\pd{r}\Pi=0$ is applied here.
When comparing with the problem for the velocity, here
the difficulty is not the coupling of the boundary 
condition for $\mathbf{B}$ with $\Pi$, 
but between the components of $\mathbf{B}$ at an insulating boundary.

Here the system is split as in (\ref{eq:NSbulk}) for the `bulk'
solution, with approximate boundary condition 
$\bar{\mathbf{B}}=\mathbf{B}^q$ on $\{r_i,r_o\}$.  This is then
corrected precisely via the influence matrix 
requiring only the simple auxiliary system
\begin{equation}
   \label{eq:inflpert}
  X \mathbf{B'} = \mathbf{0},
\end{equation}
with boundary conditions 
$B'_\pm=\{0,1\}$ or $\{1,0\}$ and
$B'_z=\{0,\mathrm{i}\}$ or $\{\mathrm{i},0\}$ 
on $\{r_i,r_o\}$.  

Problem (\ref{eq:inflpert}) separates for the 
three components, which, with the 
two boundary condition options for each,
provides six functions 
$B'_j(r)$.  
The correction is then
\begin{equation}
   \label{eq:superpos}
   \mathbf{B}^{q+1} = \bar{\mathbf{B}} + \sum_{j=1}^6 a_j\, \mathbf{B}'_j \, .
\end{equation}
Let $g_j(\mathbf{B})=0$ denote the insulating boundary conditions 
and solenoidal condition evaluated at $r_i$ and $r_o$.
Substituting (\ref{eq:superpos}) into the 
boundary conditions, they may be written
$
  A \mathbf{a} = -\mathbf{g}(\bar{\mathbf{B}}) ,
$ 
where $A=A(\mathbf{g}(\mathbf{B}'))$ 
is a 6$\times$6 matrix.  The appropriate coefficients required
to satisfy the boundary conditions are recovered from
solution of this small system for $\mathbf{a}$.

Again, the auxiliary functions $B'_j(r)$, the matrix 
$A$ and its inverse may be precomputed, and
the boundary conditions for $\mathbf{B}'$ have been chosen
so that $B'_\pm$ are purely real, 
$B'_z$ is purely imaginary, and $A$ is real.  
At the end of the timestep, the solution is solenoidal and
satisfies the boundary conditions to machine precision.

\subsection{Implementation notes and parallelization}

The Taylor-Couette flow code was written in Fortran90.  Nonlinear terms are evaluated using the pseudo-spectral method and are de-aliased using the 3/2 rule. The Fourier transforms are performed with the FFTW3 library \cite{frigo2005design} and matrix and vector operations are performed with BLAS \cite{lawson1979basic}. Each predictor-corrector iteration involves the solution of banded linear systems with forward-backward substitution using banded LU-factorizations that are precomputed prior to time-stepping. These operations are performed with LAPACK \cite{laug}. The code was parallelized so that data is split over the Fourier harmonics for the linear parts of the code: evaluating curls, gradients and matrix inversions for the time-stepping (these linear operations do not couple modes). Here all radial points for a particular mode are located on the same processor; separate modes may be located on separate processors. Data is split radially when calculating Fourier transforms and when evaluating products in real space (nonlinear term of the equations). The bulk of communication between processors occurs during the data transposes.

\subsection{Numerical validation}

The code was validated against several published linear stability results, as well as three-dimensional nonlinear simulations of the coupled induction and Navier-Stokes equations. 
We tested the inductionless limit $Pm=0$ and finite $Pm$, obtaining excellent  agreement in all cases.

\subsubsection{Linear stability of Couette flow subject to magnetic fields}

Linear instabilities were detected in the calculations
by monitoring the kinetic energy of the deviation from 
circular Couette flow after introduction of a small disturbance. 
In the linear regime we write
$$
 u' \sim \exp(\lambda t + \mathrm{i}[k z +m \phi]),
 \qquad  E \sim |u'|^2 \sim \exp(2\sigma t),
$$
where $\lambda = \sigma +\mathrm{i}\omega$ is a complex number; the imaginary part $\omega$ is the oscillation frequency and the real part $\sigma$ the growth rate of the dominant perturbation. The latter is readily extracted from the relationship $\log(E) \sim 2 \sigma t$.  

We first reproduced the classical results of Roberts \cite{roberts1964stability}, who considered the inductionless limit $Pm=0$ for narrow gap  geometry $\delta=0.95$ and stationary outer cylinder. For a Hartmann number of $Ha=5.477$ he obtained a critical Reynolds number of $Re_c=281.05$ with associated critical axial wavenumber of $\alpha_c=2.69$ and $m=0$. In our simulations we fixed $\alpha=\alpha_c$ and obtained $Re_c=281.055$ using $N=33$ radial points.  For finite magnetic Prandtl number $Pm=1$ we reproduced the results of Willis and Barenghi \cite{willis2002} for wide gap $\delta=0.5$ and stationary outer cylinder. For $Ha=39$, and $\alpha=2.4$ and $m=0$ they found $Re_{c}=60.5$, which is in good agreement with our result ($Re_{c}=60.3$). In order to test the azimuthal magnetic field we reproduced recent results  of  Hollerbach \emph{et al}.~\cite{hollerbach2010nonaxisymmetric} for the AMRI. For example at $Pm=0$, $Ha=316$, $Re=1000$, $\delta=0.5$ and $\mu=0.26$, they obtained $\sigma=-78.6$ for wavenumbers $\alpha=7.17$ and $m=1$, which is in very good agreement with our value $\sigma=-78.7$.

\subsubsection{Nonlinear simulations}

Willis and Barenghi \cite{willis2002taylor} explored dynamo action in Taylor-Couette flow. They first solved the Navier-Stokes equations in the absence of magnetic field and subsequently applied a small magnetic disturbance to test whether it grew into a dynamo. In the axisymmetric Taylor-vortex regime
axisymmetric magnetic fields were found to decay, in accordance with Cowling's anti-dynamo theorem.  Non-axisymmetric magnetic fields
may be excited, however, and for $Re =136.4$, $\delta=0.5$, $Pm=2$, $\alpha =1.57$ and stationary outer cylinder they observed that the magnetic disturbance grows for $m=1$ ($\sigma_{B,m=1}\approx 0.2$, leading to dynamo action), whereas it decays for $m=2$ ($\sigma_{B,m=2}\approx-1.4$).  We reproduced this setting using $N=41$, $K=16$ and $M=12$   and obtained $\sigma_{B,m=1}\approx 0.16$ and 
$\sigma_{B,m=1}\approx -1.42$, in good agreement with \cite{willis2002taylor}.

Finally, we compared results of the axisymmetric HMRI (helical field $\mathbf{B}= B_0 (\mathbf{e_z}+\gamma \mathbf{e_\phi})$) obtained with the spectral code of Hollerbach \cite{hollerbach2008spectral}. A typical diagnostic quantity is the torque at the cylinders
\begin{equation}
\label{eq:torque}
G \sim - 2 \pi r^3 \frac{\partial}{\partial r} \Big [\frac{u_\phi}{r} \Big ] \sim 2 \pi r^2 \Big [ \frac{u_\phi}{r} - \partial_r u_\phi \Big ].
\end{equation}
The laminar flow torque will be used as a scale, so that the 
dimensionless ratio
$G/G_\text{laminar}$ 
measures the intensity of angular momentum transfer relative to 
laminar flow. 
We choose the parameters $Re=300$, $Ha=10$, $\delta = 0.5$,  $\gamma=2$, $\alpha=0.314$, which are well into the nonlinear regime. After nonlinear saturation the dimensionless torque on the cylinders obtained with our code for  $N=81$ and $K=192$ was $G/G_\text{laminar}=1.4122$, which is in excellent agreement with the code of Hollerbach
 ($G/G_\text{laminar}=1.4123$).

\section{Primary instability: standing waves}

In the experiments of Seilmayer \emph{et al.} \cite{seilmayer2014experimental} the AMRI was explored near the onset of instability for two different Reynolds numbers $Re = 1480$ and $2960$, and Hartmann numbers in the range $Ha\in[0,160]$. The experiments have an aspect-ratio of $10$. Here we selected a periodic domain of length  $L_z=12.6$, and initialised the simulations by disturbing all Fourier modes with the same amplitude, thus allowing the axial wavenumber to be naturally selected. 
Because of the symmetries (see \S\ref{sec:sym}), two different Hopf-bifurcation scenarios are possible \cite{knobloch1996symmetry}. In the first one, the $z$-reflection symmetry is broken and depending on the initial conditions either upward traveling waves (with $k>0$ modes) or downward traveling waves (with  $k<0$ modes) may be observed. In the second scenario, the $z$-reflection symmetry is preserved and a standing wave emerges. This is a combination of upward and downward traveling waves for which positive and negative $k$ modes are in phase and have exactly the same amplitude. In both scenarios waves rotate in the azimuthal direction.

\begin{figure}
  \begin{center}
    \begin{tabular}{cc}
     (a) \hspace{1cm} & (b)\\ \includegraphics[width=0.35\linewidth]{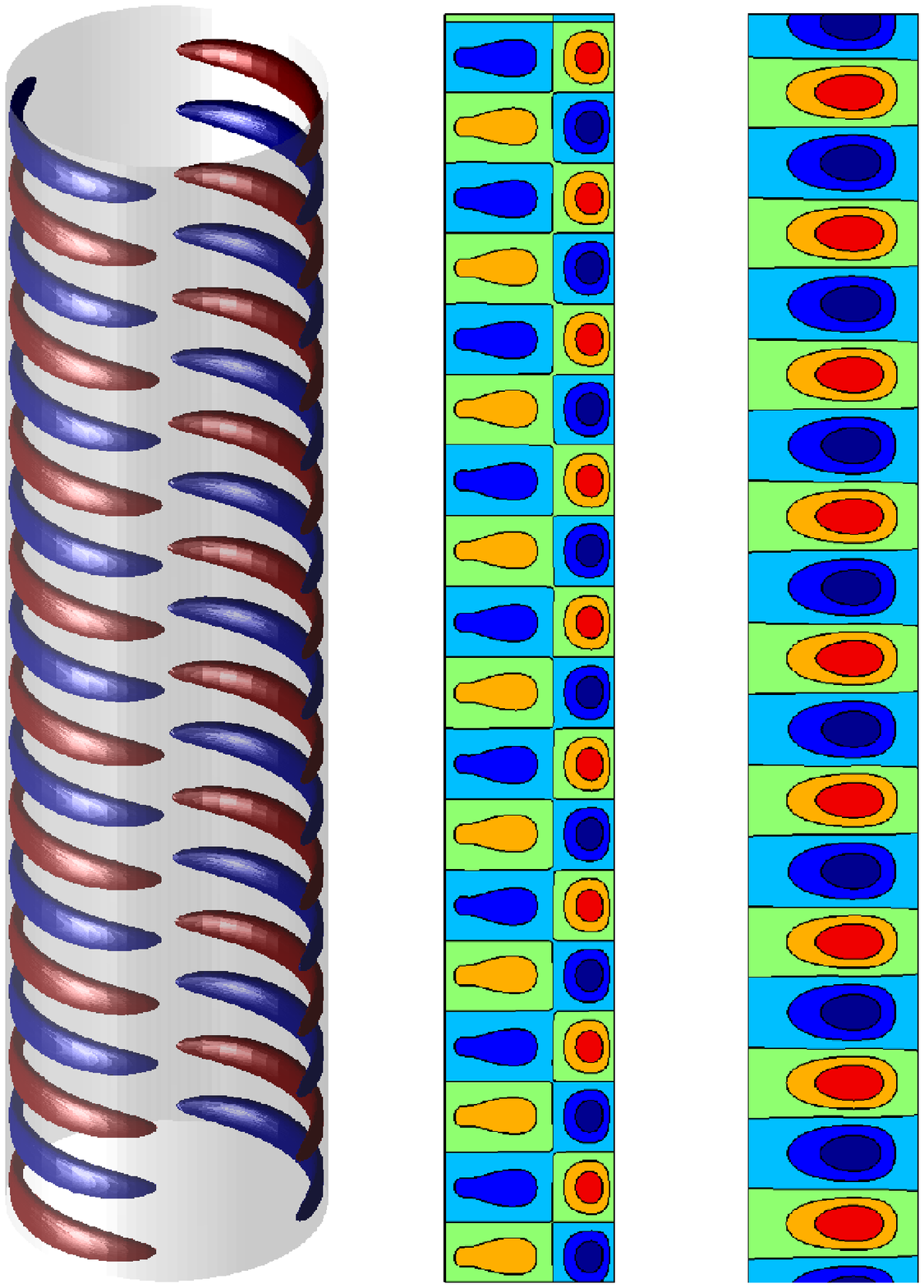} \hspace{1cm}&
     \includegraphics[width=0.45\linewidth]{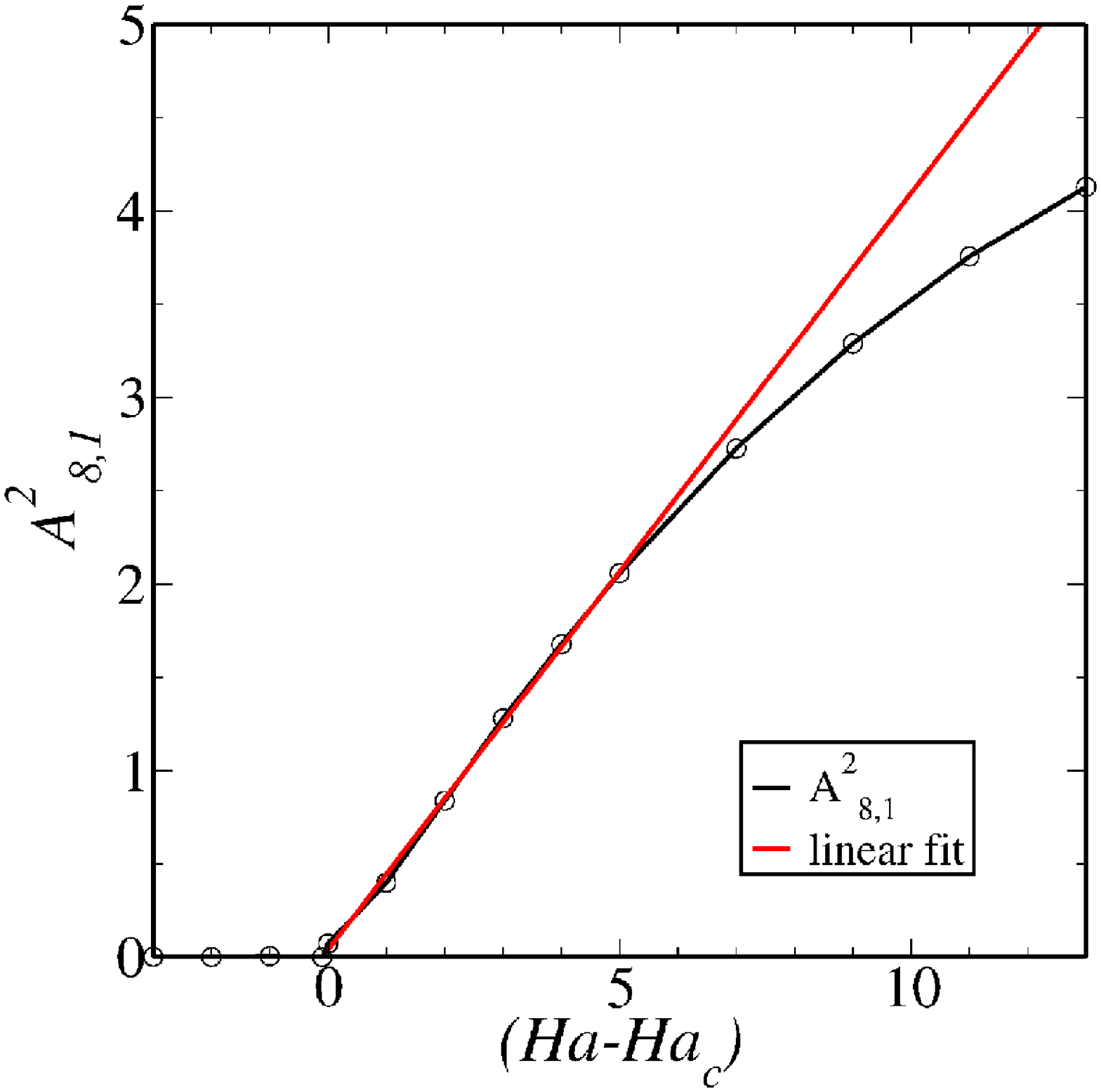}\\
    \end{tabular}
  \end{center}
  \caption{\small \textbf{Primary instability: a standing wave arises through a supercritical Hopf bifurcation}. ($a$) \textbf{Standing wave} with $k=9$ (corresponding to 9 vortex-pairs in the axial direction) at $Re=1480$, $Ha=150$ and $L_z=12.6$ (long domain). From left to right: isosurfaces of axial velocity \AG{$v_z= \pm 0.005$ (normalized with the velocity of the inner cylinder  $\Omega_i r_i$)}, contours of axial and radial velocity. \AG{The aspect ratio of the colormaps has been stretched by a factor of $0.6$.} ($b$)
   \textbf{Onset of instability} at $Re=1480$. The critical Hartmann number is $Ha_{c} \approx 107$ with critical axial mode $k=8$. The square of the amplitude of the Fourier coefficient $A_{8,1}^2$  depends linearly on $Ha - Ha_{c}$ close to the critical point as expected in a Hopf bifurcation. The coefficient $A_{-8,1}$ has the same amplitude as $A_{8,1}$, confirming that the axial reflection symmetry is preserved (standing wave).}
  \label{fig:sup_Hopf}
\end{figure}

\begin{figure}
  \begin{center}
    \begin{tabular}{cc}
     (a) & (b) \\ \includegraphics[width=0.475\linewidth]{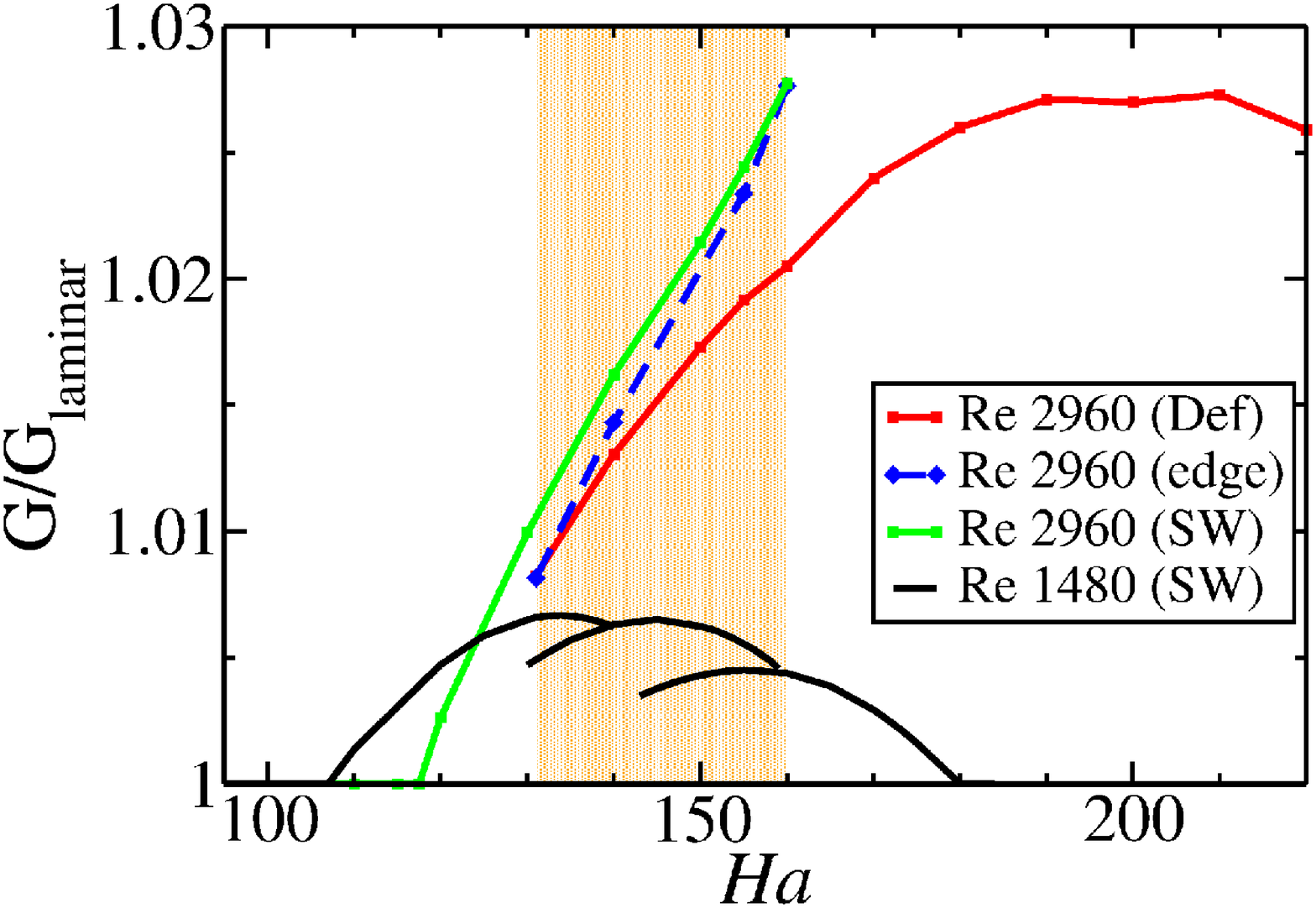}&
    \includegraphics[width=0.495\linewidth]{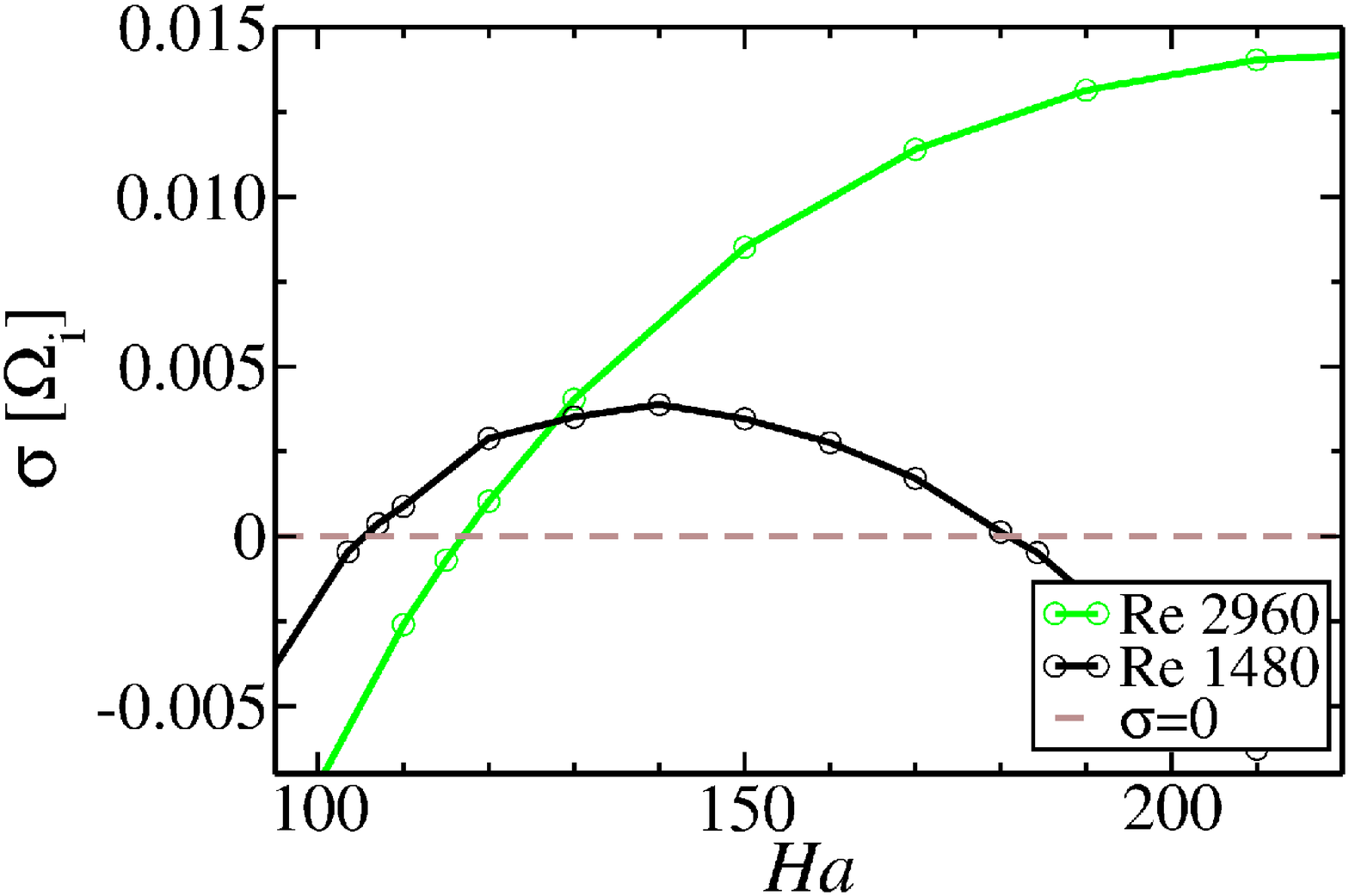}\\
    \end{tabular}
  \end{center}
  \caption{\small \textbf{Onset of spatio-temporal chaos}. ($a$) \textbf{Dimensionless torque for AMRI} versus $Ha$ for $Re = 1480$ and $Re = 2960$. Eckhaus instability at $Re=1480$: the branches of the black curve belong to different axial wavenumbers (k=8, 9 and 10) of the standing wave. Bistability at $Re = 2960$: in the yellow-shaded region standing waves (green) and defects (red) coexist; between them there is an unstable branch or edge state (blue). \AG{($b$)   \textbf{Perturbation growth rates} $\sigma$ (normalised with $\Omega_i$) as a function of $Ha$ for $Re = 1480$ and $Re = 2960$. Positive values of $\sigma$ correspond to  instability.}}
  \label{fig:subcr_Hopf}
\end{figure}

We found that at $Re=1480$ the circular Couette flow (\ref{eq:couette}) becomes unstable at $Ha_c = 107$. The emerging pattern is a standing wave (SW) with dominant mode $(k,m)=(\pm 8,1)$, so that 8 pairs of vortices fit in the domain. Figure~\ref{fig:sup_Hopf}b shows the square of the amplitude of the complex Fourier coefficient $A_{ 8,1}$ for increasing $Ha$. As expected in a Hopf bifurcation,  $A_{k,m}^2 \propto Ha-Ha_c$ near the onset of instability, and this relationship holds up to $Ha \approx 112$. The vortex arrangement of the standing wave at $Ha=150$ is shown in the flow snapshot of figure~\ref{fig:sup_Hopf}a. In this case the mode $k=9$ was naturally selected. Thus the dominant axial wavenumber depends on $Ha$ because of the Eckhaus instability, as also observed in hydrodynamic Taylor-Couette flow \cite{riecke1986stability}. The torque changes respectively with axial wavenumber (black curve on the Fig. \ref{fig:subcr_Hopf}a), so at the same parameter value states with different wavenumber and torque can be realised depending on the initial conditions. Further increasing $Ha$ the instability is gradually damped until it disappears at $Ha \approx 175$. Over the whole Hartmann range the additional torque due to the SW never exceeds 1\% of the laminar flow (see figure~\ref{fig:subcr_Hopf}a), indicating very weak transport of angular momentum. \AG{The maximum in torque correlates well with the maximum growth rate from the linear stability analysis shown in Fig. \ref{fig:subcr_Hopf}b}.

\begin{figure}
  \begin{center}
    \begin{tabular}{cc}
     (a) \hspace{2cm} & (b) \\ \includegraphics[width=0.35\linewidth]{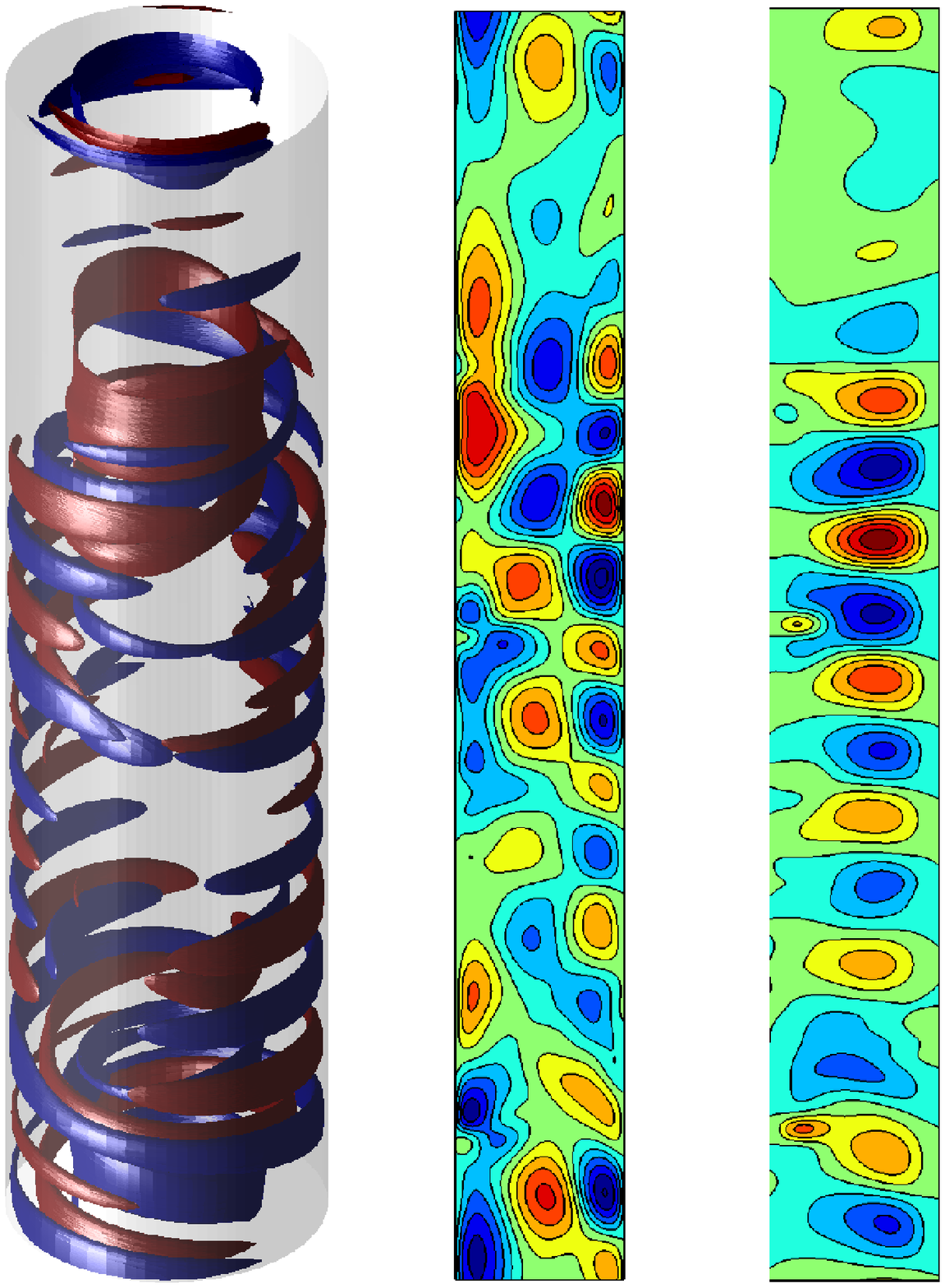} \hspace{2cm} & 
    \includegraphics[width=0.35\linewidth]{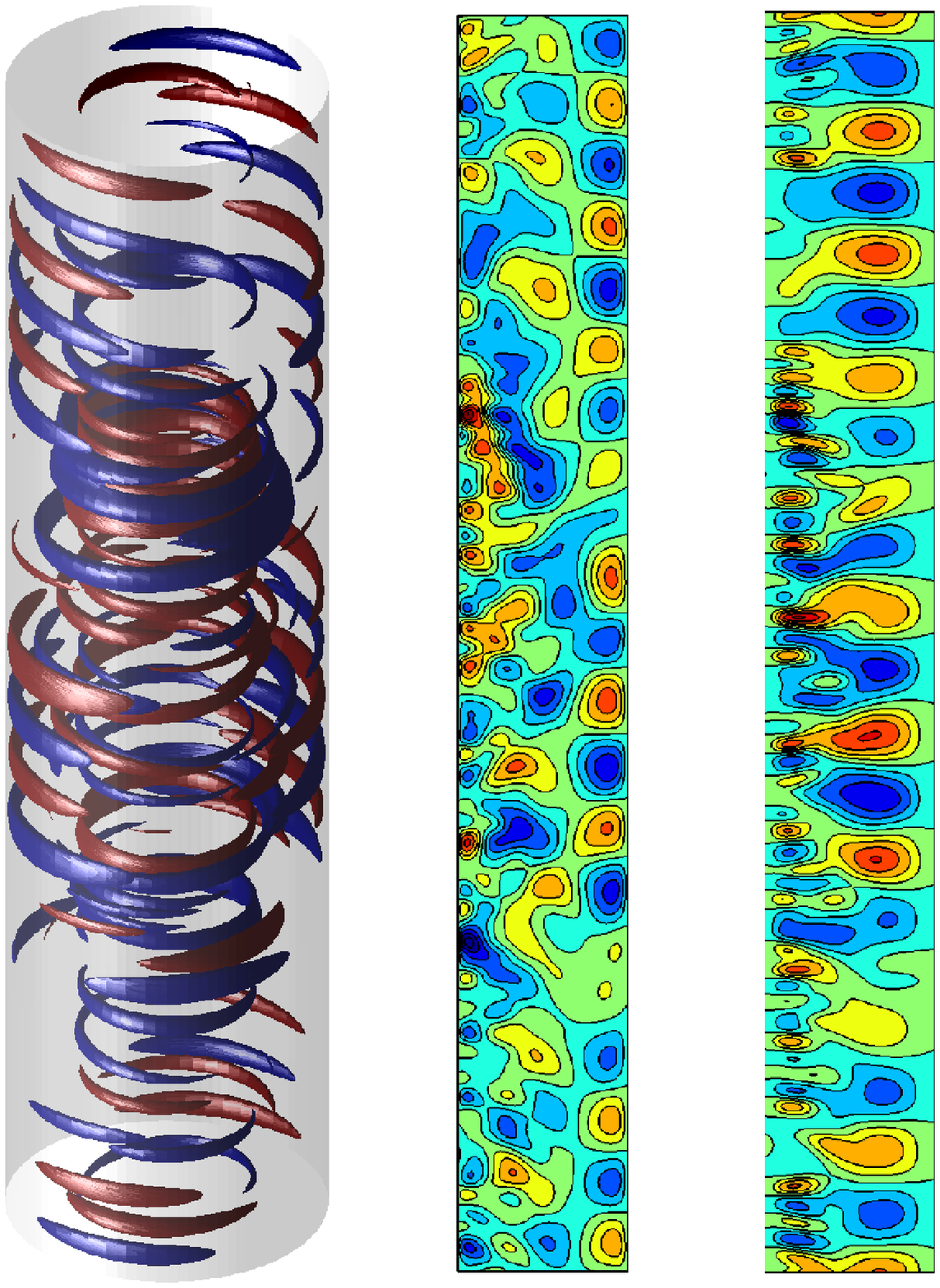}\\
    \end{tabular}
  \end{center}
  \caption{\small ($a$) \textbf{Defects at} $Re=2960, Ha=190$ and $L_z=12.6$ (long domain). From left to right: isosurfaces of axial velocity \AG{($v_z= \pm 0.01$ [$\Omega_i r_i$])}, contours of axial and radial velocity. ($b$) \textbf{ Onset of turbulence}. Isosurfaces of axial velocity \AG{($v_z= \pm 0.0125$ [$\Omega_i r_i$])}, contours of axial and radial velocity. $Re=4000$, $Ha=264$ and $L_z=12.6$. \AG{The aspect ratio of the colormaps has been stretched by a factor of $0.6$.}}
  \label{fig:defects}
\end{figure}

\begin{figure}
  \begin{center}
    \begin{tabular}{cc}
  (a) & (b) \\  \includegraphics[width=0.48\linewidth]{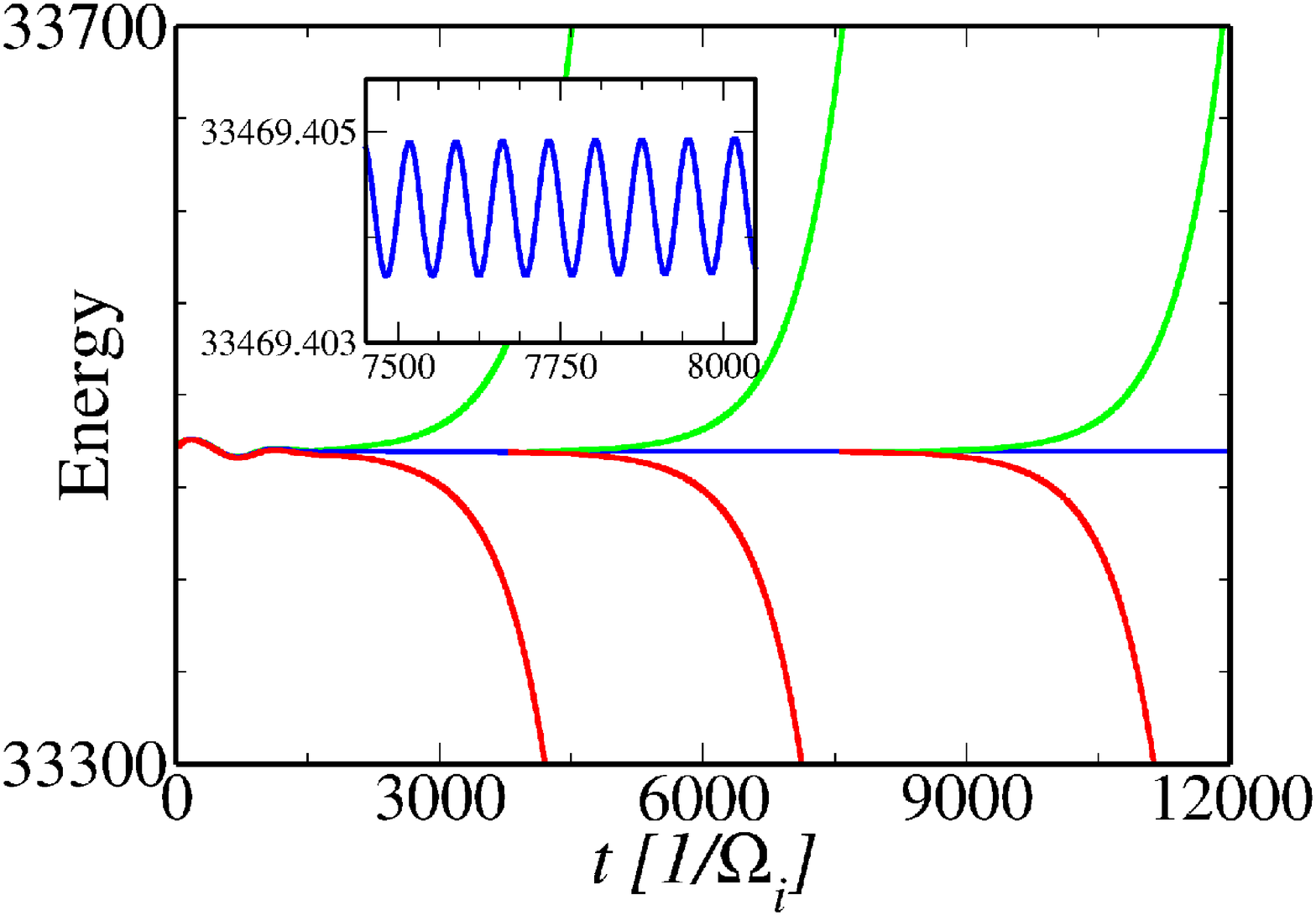}&
    \includegraphics[width=0.49\linewidth]{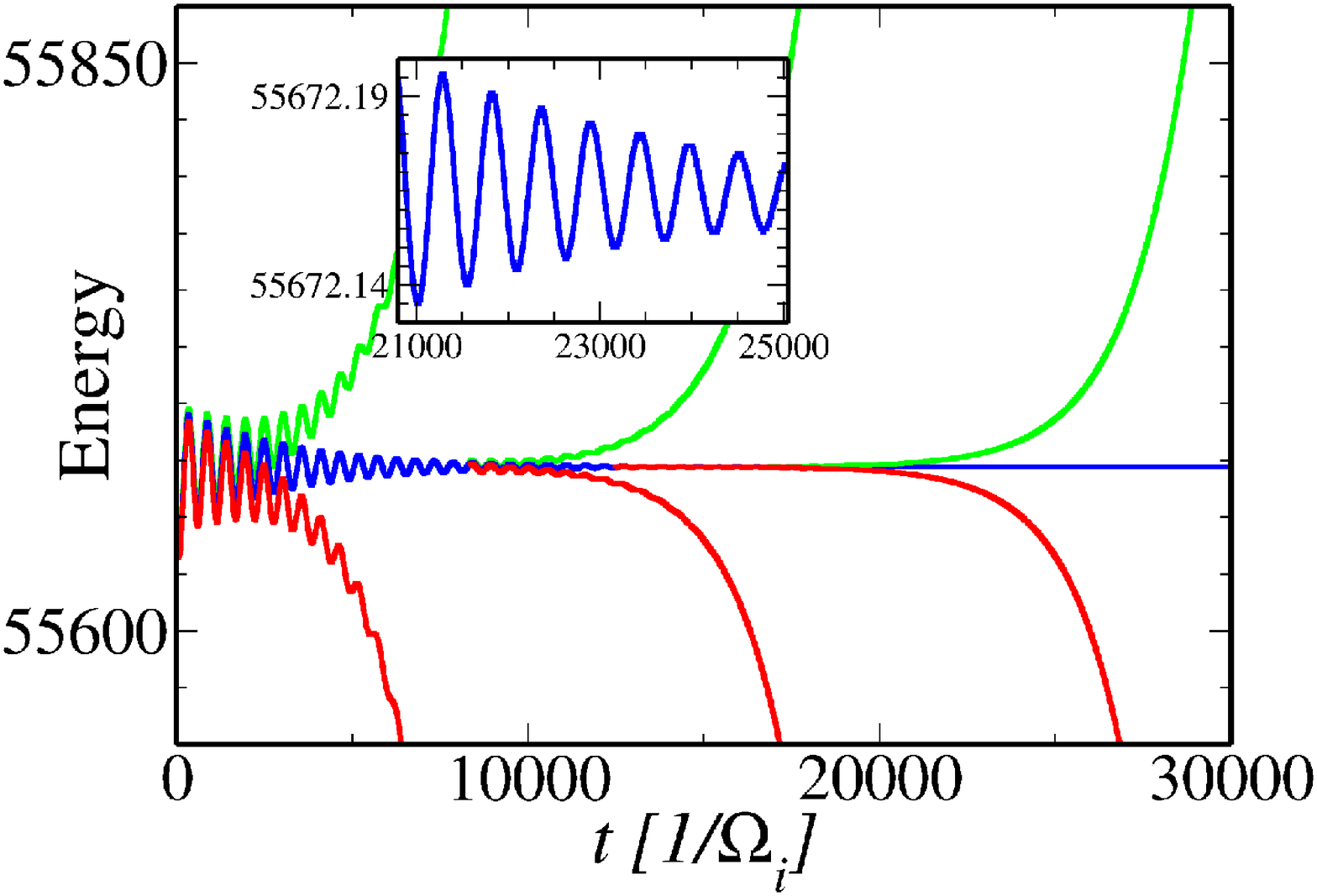}\\
    \end{tabular}
  \end{center}
  \caption{\small \textbf{Edge tracking procedure} at $Re=2960$. \AG{(a) $Ha=140$ and (b) $Ha=155$}. Green curves evolve toward the standing wave state and red curves toward defects, although all of them start very close to the edge state. \AG{Oscillations at $Ha=155$, which is close to the destabilisation point of the standing wave, appear to decay, while at $Ha=140$ they saturate. Time is normalised using the inner cylinder rotation frequency $1/\Omega_i$, i.e. $t=t \cdot Re$.}} 
  \label{fig:edge_track}
\end{figure}

\section{Onset of spatio-temporal chaos}

At $Re=2960$ a Hopf bifurcation occurs at $Ha_c = 120$ and the emerging SW remains stable until $Ha=160$. Increasing $Ha$ beyond this point a catastrophic transition to spatio-temporal chaos is observed: the vortex structure is damaged and the up-down symmetry is broken (Fig. \ref{fig:defects}). Between $Ha=130$ and $160$ there is a hysteresis region in which both SW and spatio-temporal chaos (defects)  are locally stable (see Fig. \ref{fig:subcr_Hopf}a). In this $Ha$-range, if the initial condition is a SW from another run with slightly different $Ha$, this remains stable. However, when starting for example from a randomly disturbed Couette profile the flow evolves directly to defects. 

This catastrophic transition \AG{suggests} a subcritical bifurcation. We investigated this hypothesis by computing the unstable branch separating defects and SW. For this purpose we combined time-stepping with a bisection strategy as follows. If the SW is slightly disturbed, then the flow should rapidly converge to the SW because it is locally stable. The same applies to defects. For intermediate initial conditions the flow should take a long time before asymptotically reaching either the SW or the defects. Such initial conditions were generated here by performing a linear combination between two selected flow snapshots of SW and  defects. This combination was parametrised with a variable $\beta$, for which $\beta=0$ corresponds to SW and $\beta=1$ to defects. With the bisection procedure, refining $\beta$ results in an initial condition successively closer to the manifold (or edge) delimiting the two basin boundaries. The edge is comprised of those initial conditions that tend neither to defects nor to SW, and the attractor in this manifold is referred to as an edge state \cite{skufca2006edge}. 

\begin{figure}
  \begin{center}
    \begin{tabular}{ccc}
     (a) \hspace{2cm} & (b) & \hspace{2cm} (c) \\
    \includegraphics[width=0.22\linewidth]{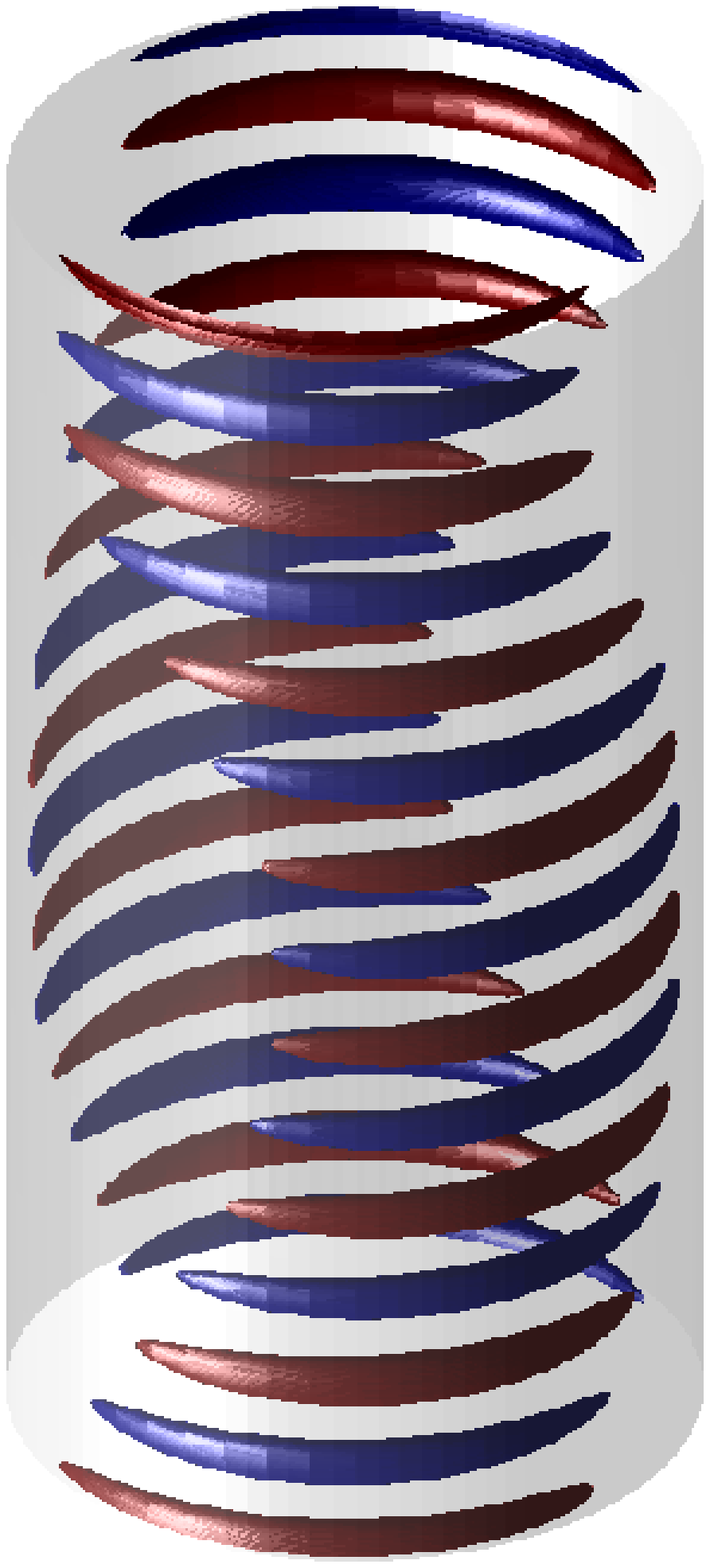} \hspace{2cm} & 
    \includegraphics[width=0.22\linewidth]{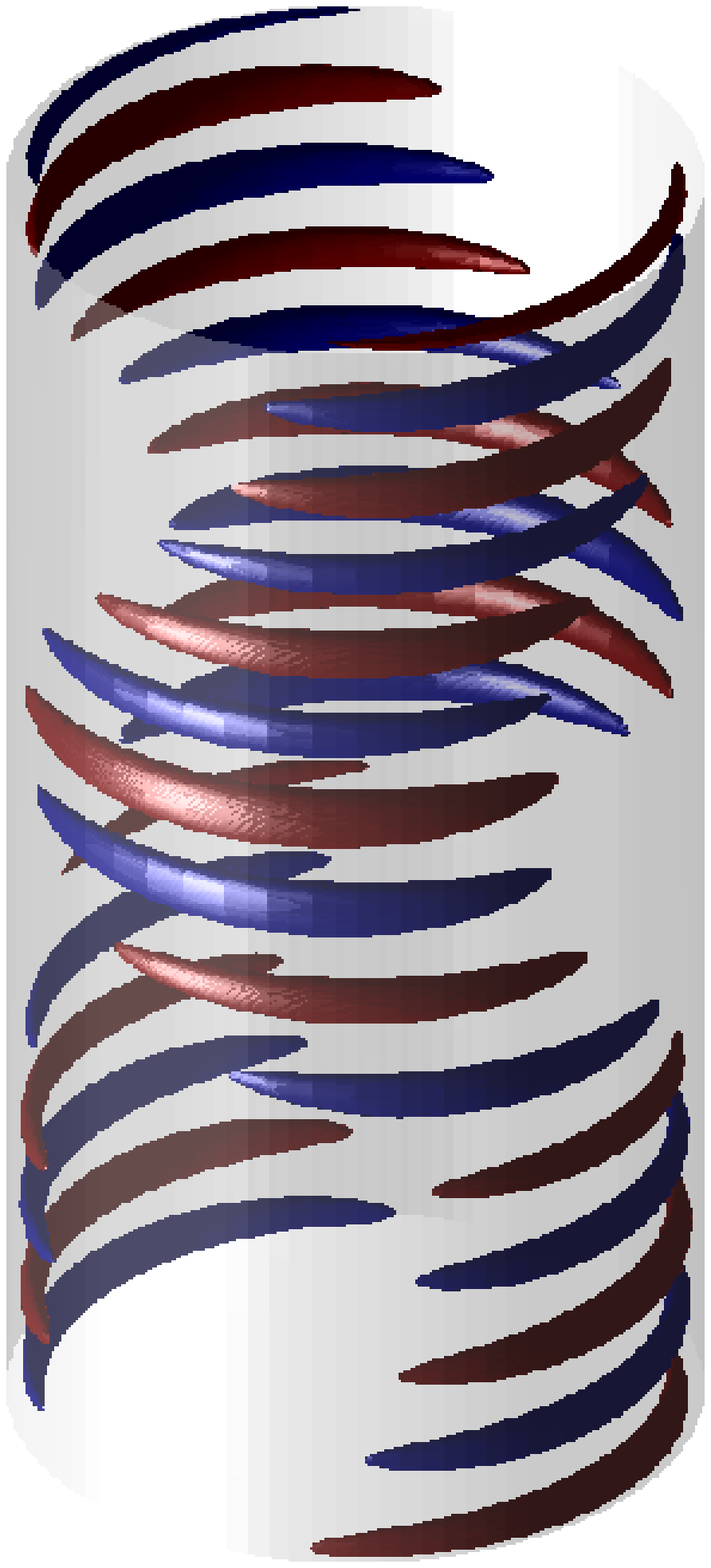} & \hspace{2cm}
    \includegraphics[width=0.17\linewidth]{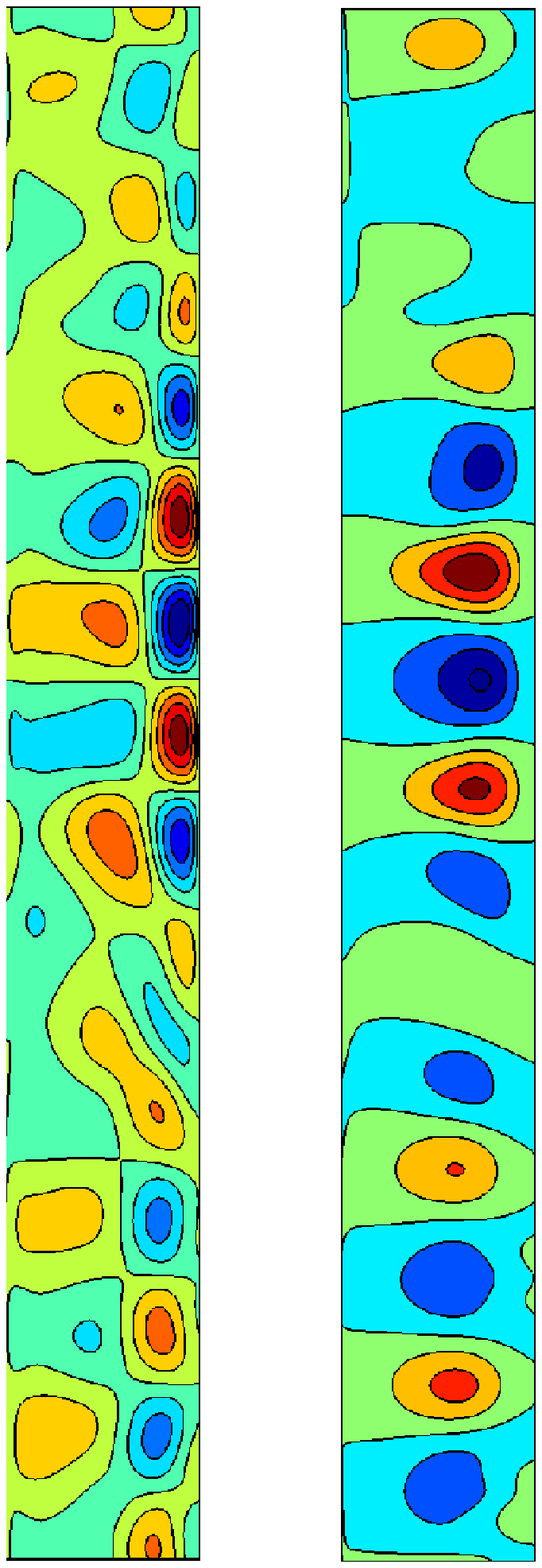}\\
    \end{tabular}
  \end{center}
  \caption{\small \textbf{Edge state at $Re=2960$ and $L_z=12.6$. (a) Close to the bifurcation point},  $Ha=155$: isosurfaces of axial velocity \AG{$v_z= \pm 0.0135$ [$\Omega_i r_i$]}. \textbf{(b) Far from the bifurcation point}, $Ha=140$: isosurfaces \AG{$v_z= \pm 0.012$ [$\Omega_i r_i$]}. \textbf{(c) Contours of axial and radial velocity}, $Ha=140$. \AG{The aspect ratio of the colormaps has been stretched by a factor of $0.6$.} The edge state consists of a long-wave modulation of the standing wave. }
  \label{fig:edge_iso}
\end{figure}

Figure \ref{fig:edge_track} shows that as initial conditions are taken closer to the edge, the temporal dynamics become simple as the edge state is approached. \AG{At $Ha=155$, which is very close to the destabilisation of the SW, the dynamics appear to exhibit a damped oscillation (see Fig. \ref{fig:edge_track}b). Unfortunately, it is difficult to establish whether the oscillation finally decays or saturates at a tiny amplitude, as expected close to the bifurcation point. At $Ha=140$, however, which is further from the bifurcation point, the oscillation saturates at non-zero amplitude (see Fig. \ref{fig:edge_track}a). This suggests that the edge state is a relative periodic orbit (or modulated wave) emerging at a subcritical Hopf bifurcation of the SW. Despite this simple temporal behaviour, the spatial structure of the edge state is complicated (see Fig.~\ref{fig:edge_iso}a--c). It consists of a long-wave (subharmonic) modulation of the axially periodic pattern of the SW, which can be seen as a precursor to defects (compare Fig.~\ref{fig:edge_iso}a--b to Fig.~\ref{fig:defects}a). We expect that as $Ha$ is further reduced the edge state suffers a bifurcation cascade and becomes chaotic. This should continuously connect to defects and stabilise at a turning point for $Ha\gtrsim130$, which is the lowest $Ha$ for which defects remain stable.}

\section{Turbulent transport of momentum}

As the Reynolds number is further increased, defects are expected to grow gradually into turbulence. \AG{Although it would be very interesting to perform a two-parameter study of the dynamics in $Ha$ and $Re$, this is computationally expensive and beyond the scope of the current work. We here chose to follow a parameter path of the form
\begin{equation}\label{eq:path}
Ha=a \,Re^b,
\end{equation}
with $a=0.71$ and $b=1.55$. This path is shown as a solid green line in Fig.~\ref{fig:TC_scheme}b and provides a very good approximation to the curve of maximum growth rate of the linear stability analysis (red dashed line). It goes deep into the instability region and so we expect the instability to fully develop as $Re$ increases with $Ha$ subject to \eqref{eq:path}.} 

\begin{figure}
  \begin{center}
    \begin{tabular}{cc}
     (a) & (b) \\ \includegraphics[width=0.5\linewidth]{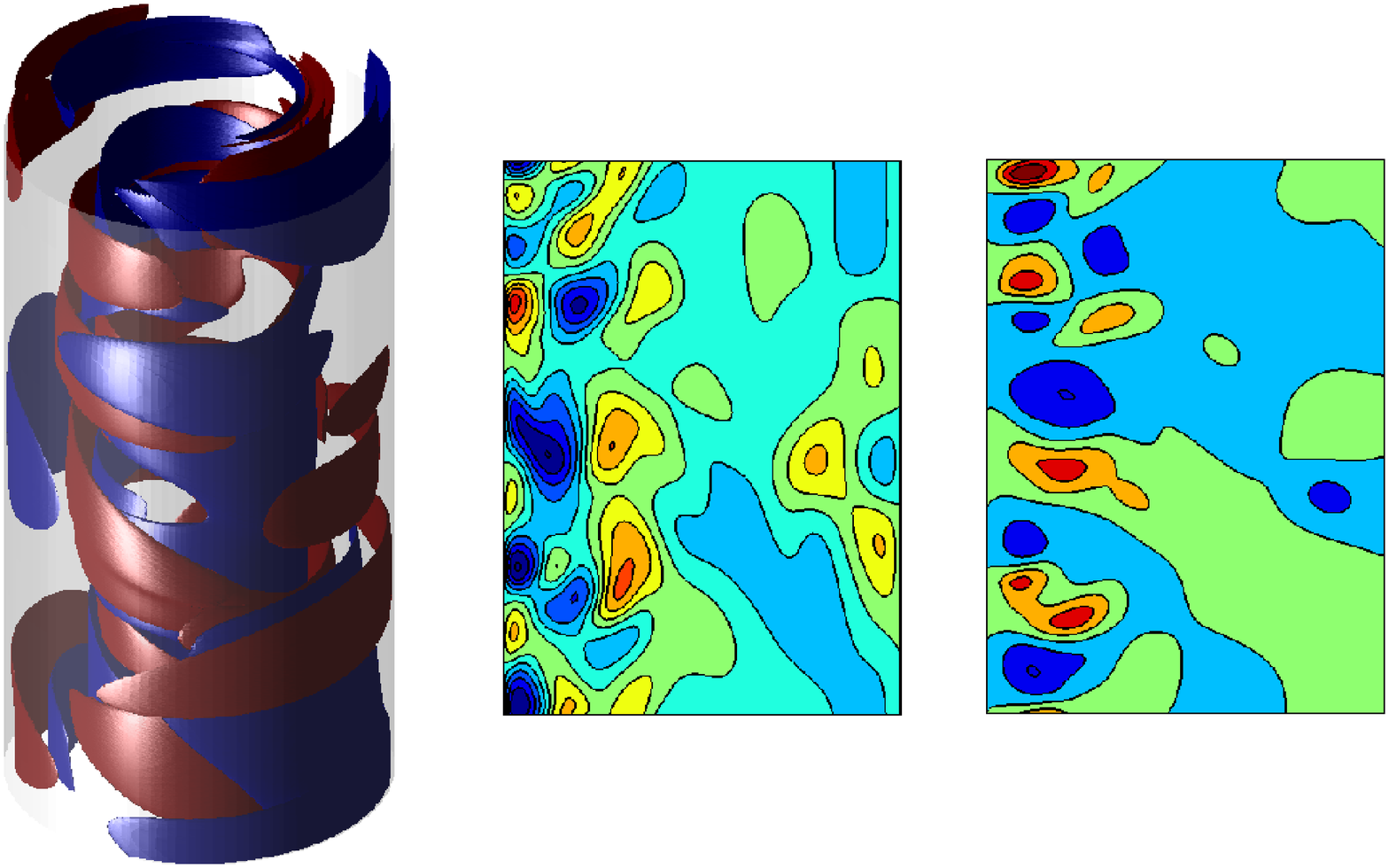} & 
    \includegraphics[width=0.45\linewidth]{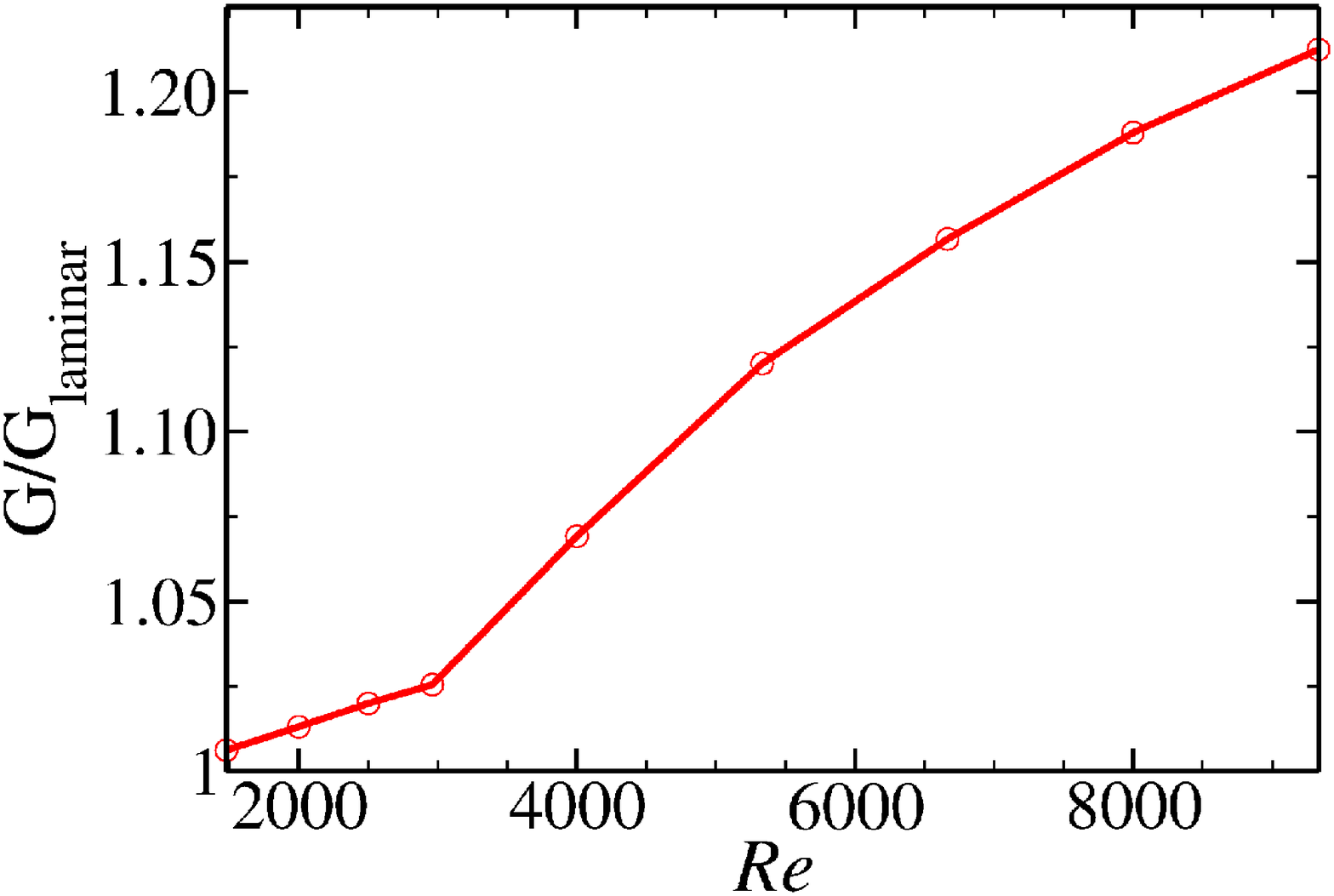}\\
    \end{tabular}
  \end{center}
  \caption{\small ($a$) \textbf{Turbulent flow in a short domain} at $Re=9333$, $Ha=456.7$ and $L_z=1.4$. Axial velocity isosurfaces \AG{$v_z= \pm 0.011$ [$\Omega_i r_i$]}, contours of axial and radial velocity. ($b$) \textbf{Dimensionless torque} as a function of $Re$ \AG{along the parameter path $Ha=a \,Re^b$, with $a=0.71$ and $b=1.55$. This path is shown as a green line in Fig.~\ref{fig:TC_scheme}b.}}
  \label{fig:turbulence}
\end{figure}

At $Re=4000$ the vortices are small at the inner cylinder and remain quite large at the outer cylinder (Fig.~\ref{fig:defects}b), and at $Re=9333$ this tendency develops into rapidly drifting small vortices at the inner cylinder and slow large vortices at the outer cylinder (Fig. \ref{fig:turbulence}a). There is no preferred direction in the system; vortices can travel up or down, both at the inner and outer cylinders.

\begin{figure}
  \begin{center}
    \begin{tabular}{cc}
     (a) & (b) \\ \includegraphics[width=0.47\linewidth]{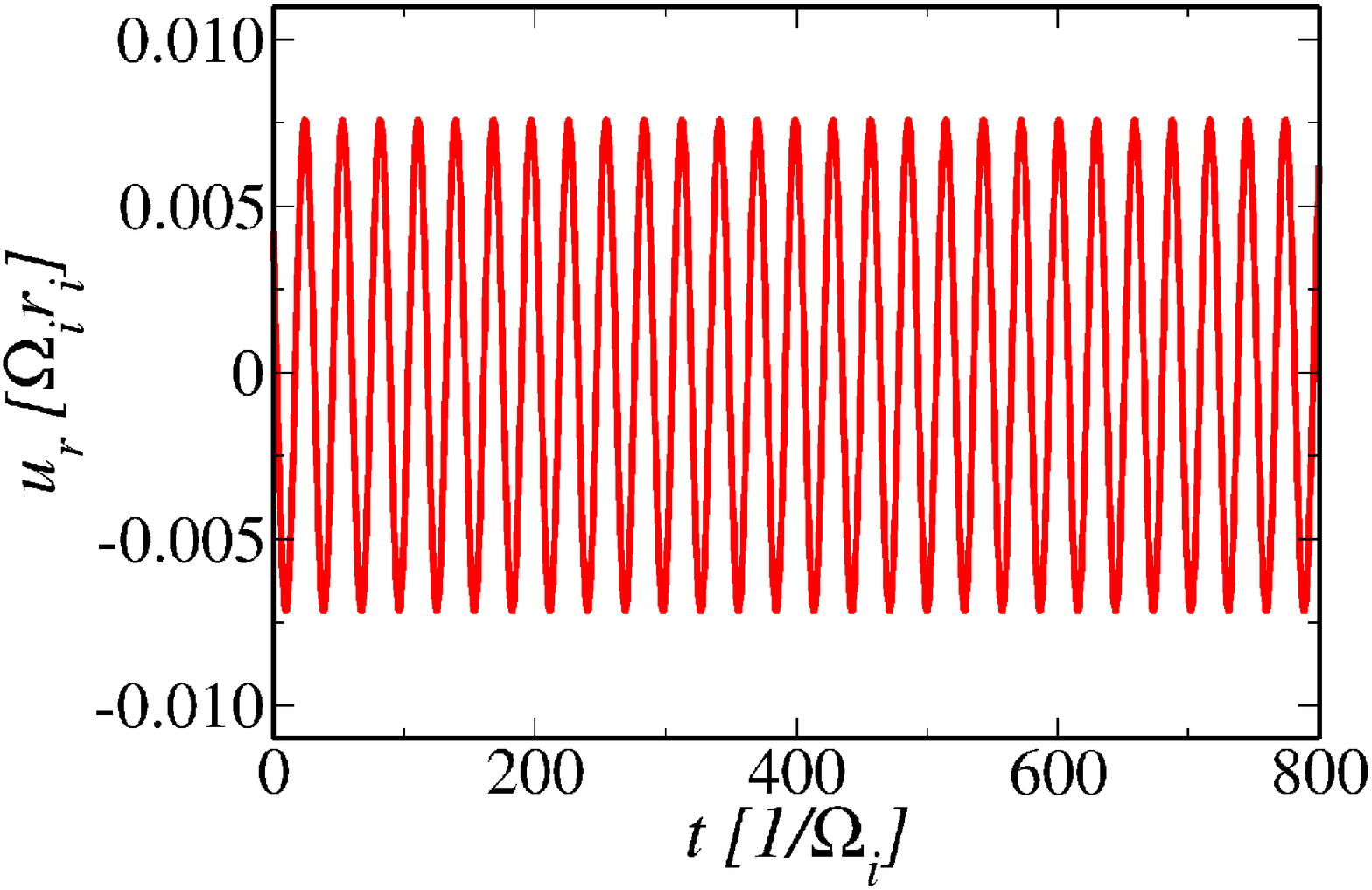}  \hspace{1cm}& 
    \includegraphics[width=0.47\linewidth]{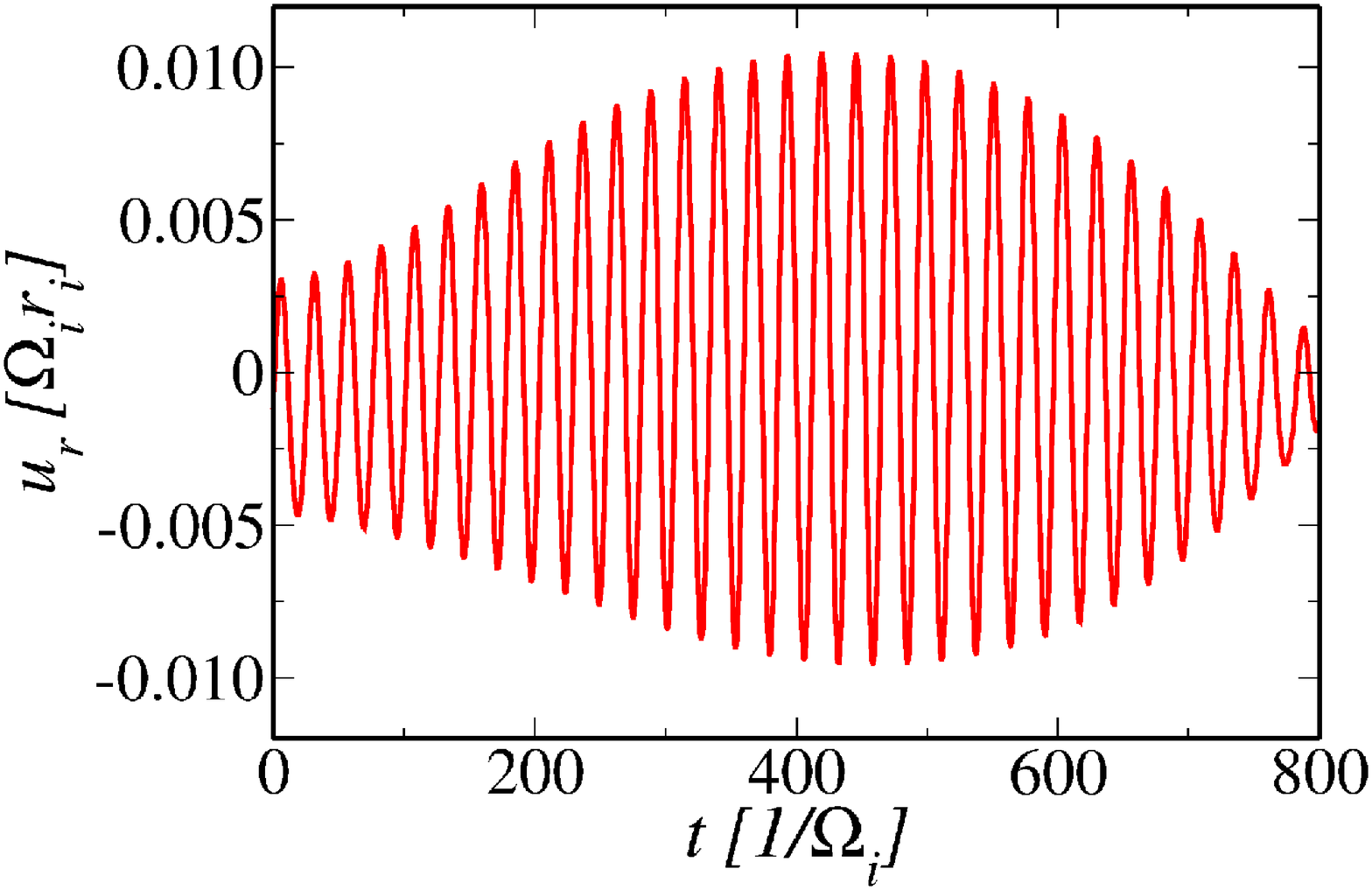}\\
     (c) & (d) \\ \includegraphics[width=0.47\linewidth]{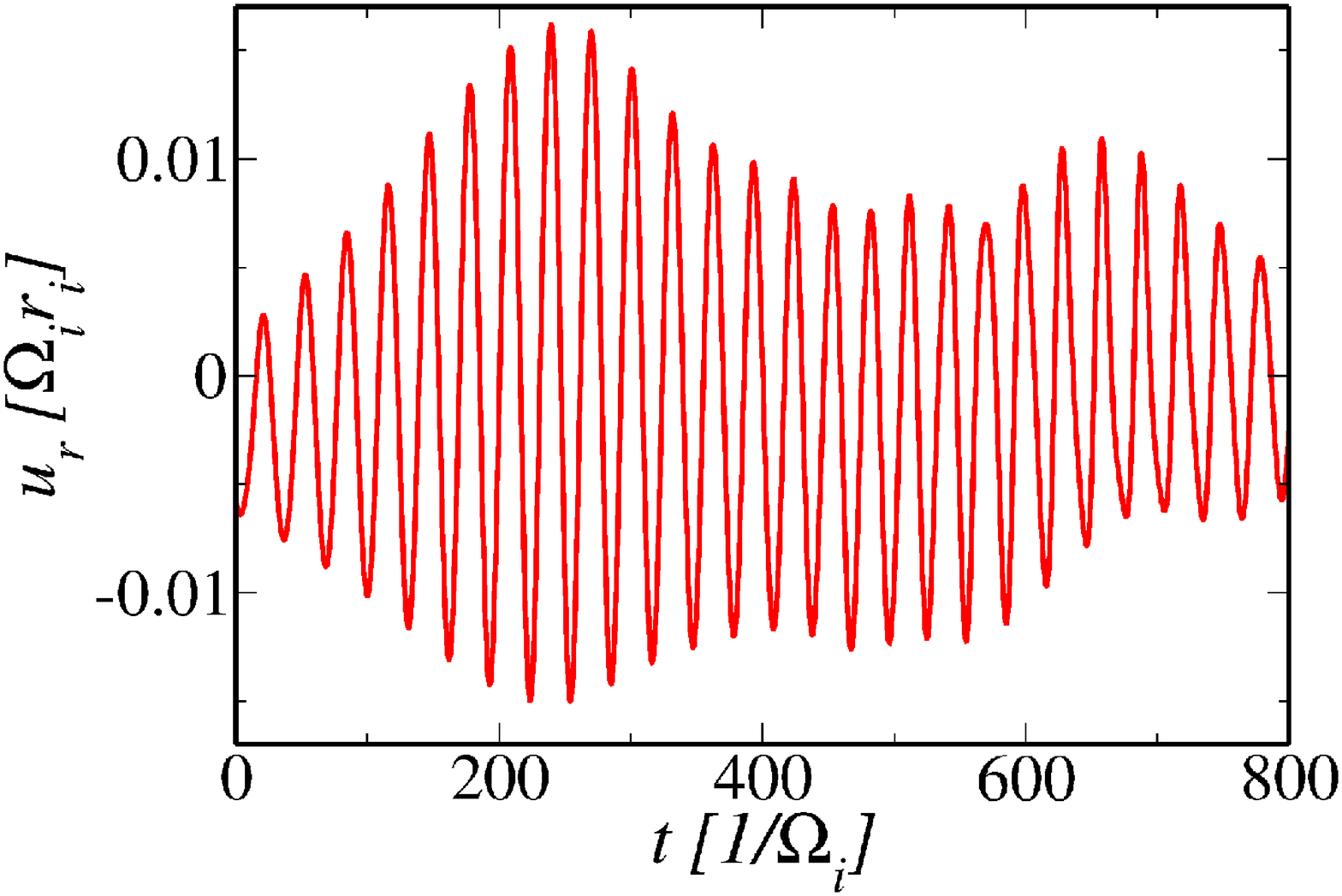}  \hspace{1cm}& 
    \includegraphics[width=0.47\linewidth]{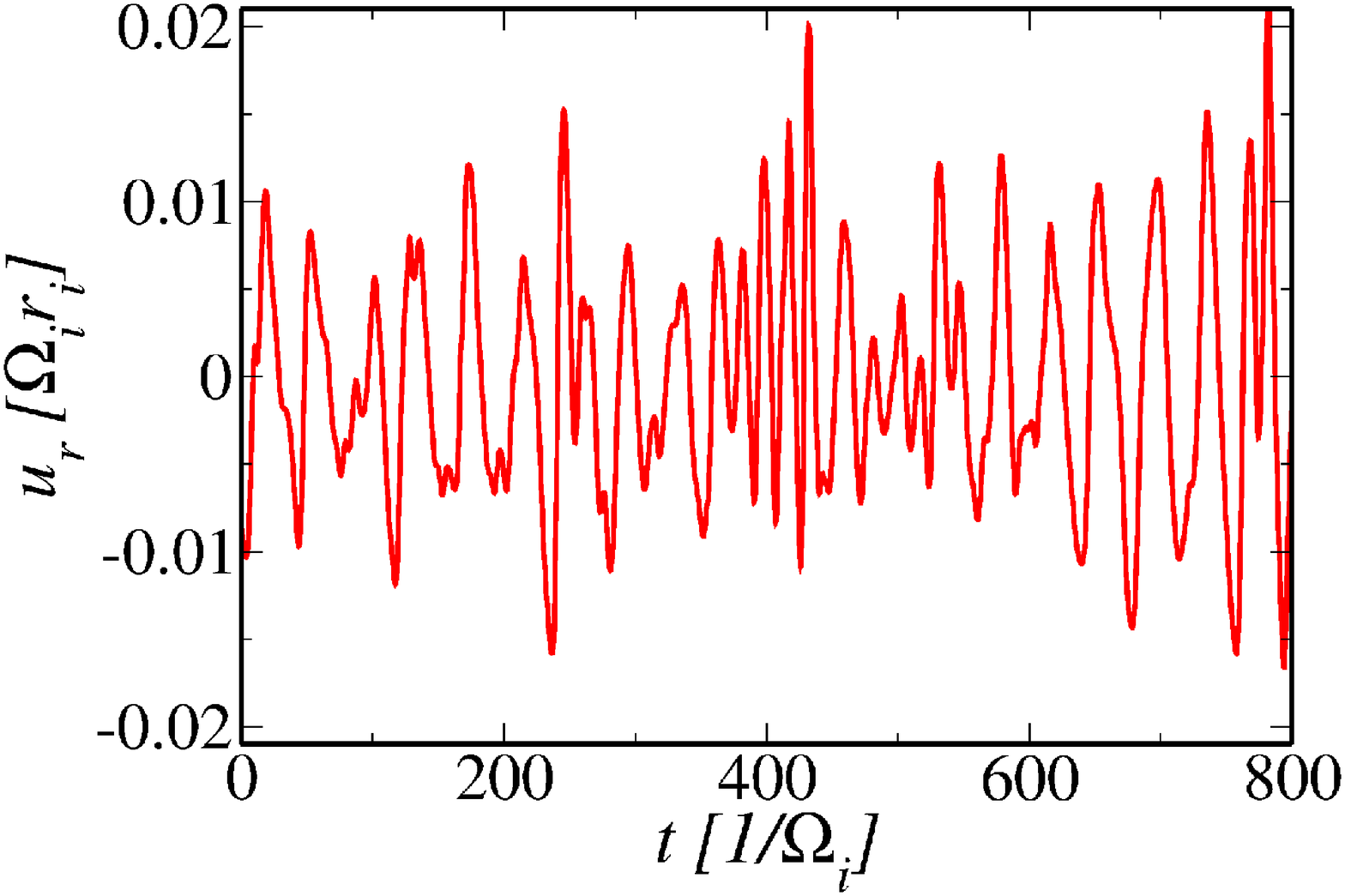}\\
    \end{tabular}
  \end{center}
  \caption{\small \textbf{Transition to turbulence.} Evolution of radial velocity perturbation $u_r$ at the point $(r, \phi, z) = (1.5, 0, 0)$ with time ($a$) at $Re = 1480$ and $Ha=150$ (standing wave); ($b$) - the same at $Re = 2960$ and \AG{$Ha =140$} (edge state); ($c$) -  at $Re = 2960$ and $Ha=190$ (defects); ($d$) - at $Re = 9333$ and $Ha=456.7$ (turbulence). Time is scaled using the inner cylinder rotation frequency $1/\Omega_i$, i.e. $t=t \cdot Re$; \AG{velocities are normalised with the velocity of the inner cylinder $\Omega_i r_i$}.}
  \label{fig:tran_turb}
\end{figure}

The qualitative difference between standing wave, defects and turbulent flow is apparent in time series of the radial velocity $v_r$ taken at the mid-gap between the cylinders $(r, \phi, z) = (1.5, 0, 0)$.
\AG{Figure~\ref{fig:tran_turb}a shows that the radial velocity of the standing wave oscillates periodically around zero. The edge state features a slow temporal frequency modulating the oscillation of the SW (Fig. \ref{fig:tran_turb}b). For defects at $Re=2960$ the time series is mildly chaotic. As $Re$ increases toward turbulence the velocity pulsates in a very chaotic manner (Fig. \ref{fig:tran_turb}d). However, the main frequency associated with the AMRI can still be discerned. By comparing all panels it becomes apparent that this frequency scales with the rotation-rate of the inner cylinder. This is consistent with the linear stability analysis of \cite{hollerbach2010nonaxisymmetric}, and with the studies \cite{kirillov2010,kirillov2012}, where it is shown that in the low $Pm$ limit the AMRI is an inertial wave.}

The transfer-rate of angular momentum between the cylinders is important for accretion-disc modelling. We checked the torque scaling for increasing $Re$ numbers (see Fig. \ref{fig:turbulence}b) along the parameter path~\eqref{eq:path}. The dimensionless torque increases with $Re$ according to  the scaling law 
\begin{equation}\label{eq:torq}
G \sim Re^{1.15},
\end{equation}
which is surprisingly low compared to hydrodynamic experiments in the Rayleigh unstable regime \cite{lathrop1992turbulent}.  \AG{We believe that this torque scaling comes close to being an upper bound for the torque scaling of the AMRI in the $Re$ range studied here because the maximum in torque correlates  well with maximum growth rates at low $Re$ (see Fig. \ref{fig:subcr_Hopf}). However, we must caution that in hydrodynamic Taylor--Couette flow different maxima of the torque have been observed as a function of the relative rotation of the cylinders~\cite{brauckmann2015momentum}. At large $Re$ these are not correlated to the maximum growth rate of the primary instability. Similar phenomena may occur for the AMRI.} 

\section{Discussion}

\begin{figure}
  \begin{center}
    \begin{tabular}{cc}
     (a) & (b) \\ \includegraphics[width=0.48\linewidth]{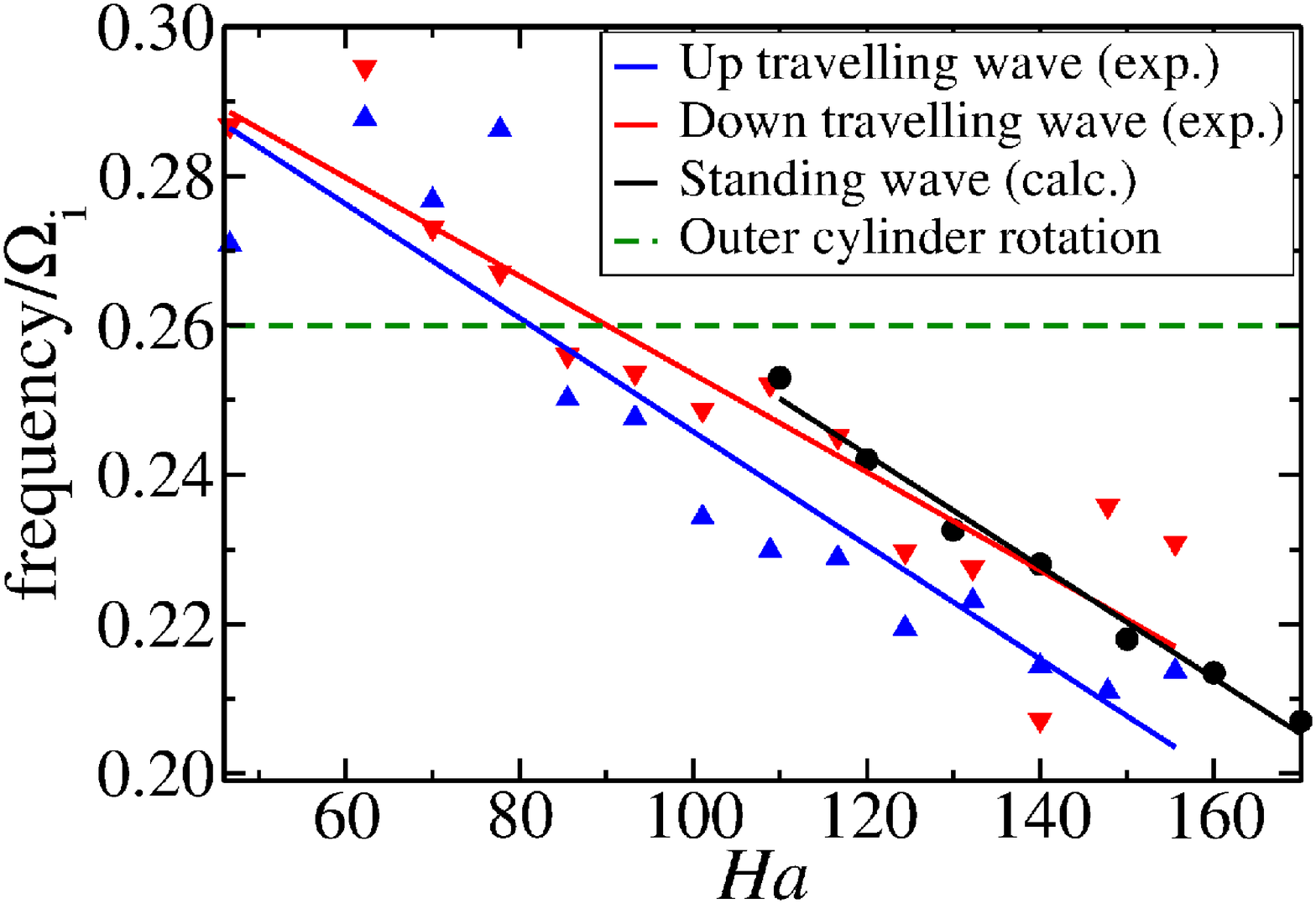} & 
    \includegraphics[width=0.48\linewidth]{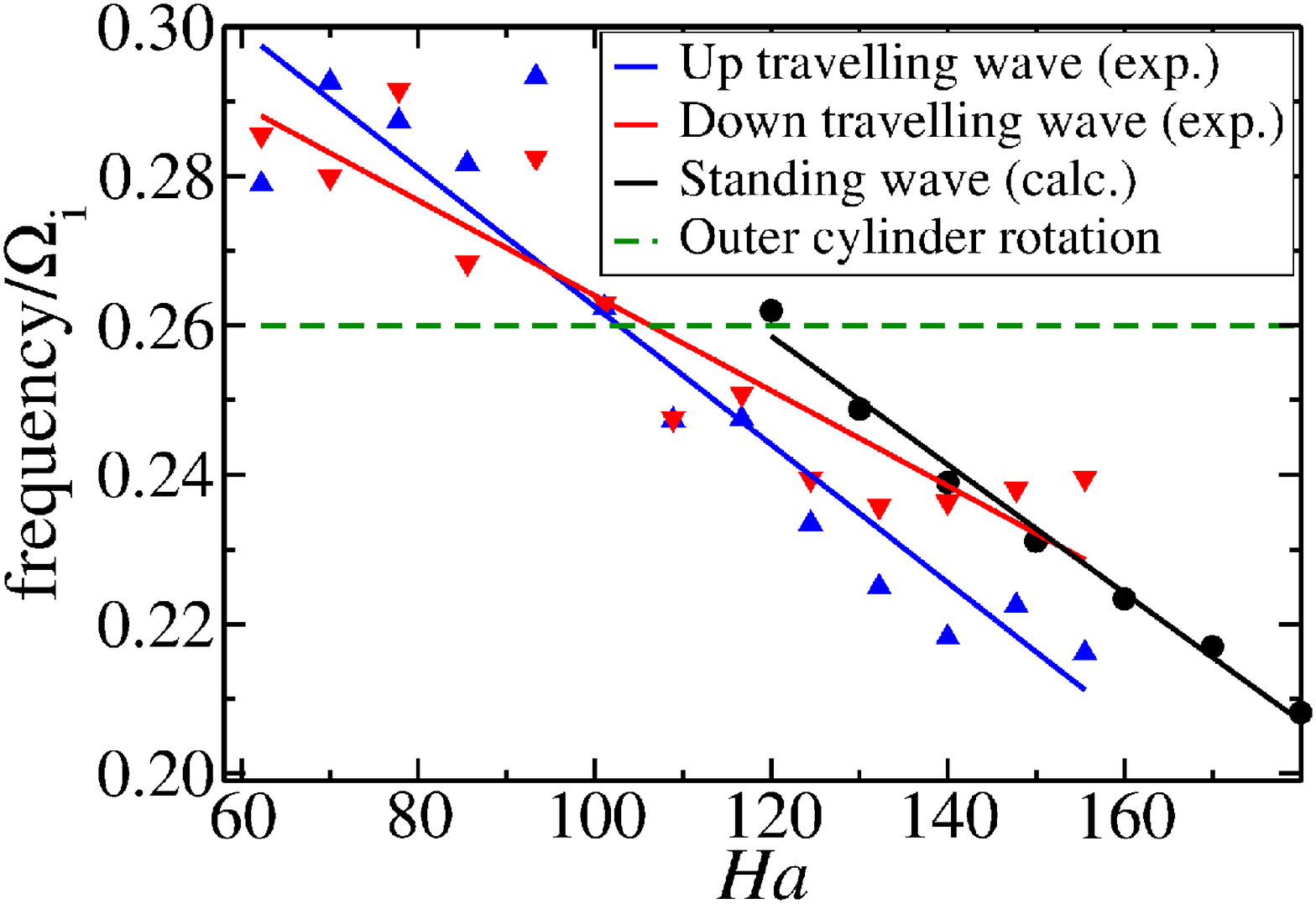}\\
    \end{tabular}
  \end{center}
  \caption{\small \textbf{Comparison to PROMISE experiment:} angular drift frequencies of the waves at ($a$) Re 1480 and ($b$) Re 2960. Blue and red lines correspond to experimental results, black to our nonlinear simulations; the green line denotes outer cylinder rotation $\Omega_o/\Omega_i=0.26$. The waves rotate at approximately the outer cylinder frequency and slow down with increasing $Ha$.}
  \label{fig:freq}
\end{figure}

We showed that the AMRI in Taylor-Couette flow manifests itself as a wave rotating in the azimuthal direction and standing in the axial direction, thereby preserving the reflection symmetry in the latter. In order to compare to experimental observations~\cite{seilmayer2014experimental} we computed the angular drift frequency of the wave. This is shown in figure~\ref{fig:freq}  after being normalised with the rotation frequency of the inner cylinder.  The wave rotates at approximately the outer cylinder frequency (dashed line) and slows down as the Hartmann number increases, which is in qualitative agreement with the experimental data. Note, however, that in the experiment two frequencies are simultaneously measured, corresponding to the up- and down-traveling spiral waves, respectively. Although in the standing wave the two frequencies are identical, in the experiment the asymmetric wiring creates $B_r$ and $B_z$ components of magnetic field which break the reflectional symmetry.  As a result, up and down spirals travel with different frequencies, similar to 
co-rotating Taylor-Couette flow in which the reflection symmetry is broken by an imposed axial flow \cite{avila2006double}. Another difference is that in the experiment the flow becomes unstable at lower $Ha$, which may be explained 
by the different boundary conditions in the experiment from our simulations. In the experiment copper cylinders are used, so perfectly conducting walls would be a closer boundary condition for the magnetic field. 
More significantly, in the experiments the cylinders are of finite length, so to reproduce their results exactly a no-slip condition on end-plates should be used. We have applied periodic boundary conditions in the axial direction, which more accurately model the accretion disc problem and allow us to compute high Reynolds number flows more efficiently. 

As Re increases, a catastrophic transition to spatio-temporal chaos occurs directly from the SW. In a range of parameters SW and chaos are both locally stable and can be realised depending on the initial conditions. We have shown that the first step in this transition process is a subcritical Hopf bifurcation giving rise to an unstable relative periodic orbit, which 
has been computed using an analogue of the edge-tracking algorithm introduced by Skufca \emph{et al}.~\cite{skufca2006edge} in shear flows. This unstable relative periodic orbit consists of a long-wave modulation of the axially periodic pattern of the standing wave and destroys the homogeneity of the vortical pattern. It can thus be seen as a temporally simple defect precursor of the ensuing spatio-temporal chaos. Because of the computational cost we could not track further instabilities on the unstable branch, which we speculate result in chaotic flow before the dynamics stabilize at a turning point ($Ha=130$ at $Re=2960$). After the turning point defects are stable and can be computed simply by time-stepping. 

We believe that such long-wave instabilities are ubiquitous in fluid flows. In linearly stable shear flows, such instabilities of traveling waves were found to be responsible for spatial localisation \cite{melnikov2014long,chantry2014genesis}. In fact, in pipe flow the ensuing localised solutions, which are also relative periodic orbits, suffer a bifurcation cascade leading to chaos \cite{avila2013streamwise}. One difference is that in pipe flow the traveling waves are disconnected from laminar flow, where the standing wave of the AMRI is connected to the circular Couette flow. 

Our simulations were performed with a powerful spectral DNS method, which we have developed and validated with published results to excellent agreement. The method allowed us to compute flows up to $Re=10^4$. As $Re$ increases, defects accumulate and the flow becomes gradually turbulent. Although we found that the AMRI exhibits a weak scaling of angular momentum transport, with $G \propto Re^{1.15}$, we expect that larger magnetic Prandtl numbers $Pm$ (realistic for accretion discs) may result in a stronger scaling. Astrophysically important issues such as the precise angular momentum transport scalings obtained for different values of $Pm$, and also for different choices of imposed field (e.g.\ SMRI, HMRI, AMRI) will be the subject of future investigations. 

\begin{acknowledgements}
Support from the Deutsche Forschungsgemeinschaft (grant number AV 120/1-1) and computing time from the J{\"u}lich Supercomputing Centre (grant number HER22) and Regionales Rechenzentrum Erlangen (RRZE) are acknowledged.  
\end{acknowledgements}

\section*{References}
\bibliographystyle{ieeetr}
\bibliography{Bibliography}

\end{document}